\documentclass[pdflatex,sn-nature]{sn-jnl}

\usepackage{fourier} 
\usepackage{graphicx}%
\usepackage{multirow}%
\usepackage{amsmath,amssymb,amsfonts}%
\usepackage{amsthm}%
\usepackage{mathrsfs}%
\usepackage[title]{appendix}%
\usepackage{xcolor}%
\usepackage{textcomp}%
\usepackage{manyfoot}%
\usepackage{booktabs}%
\usepackage{algorithm}%
\usepackage{algorithmicx}%
\usepackage{algpseudocode}%
\usepackage{listings}%
\usepackage{siunitx}
\usepackage{gensymb}
\usepackage{mathtools}
\usepackage[version=4]{mhchem}

\usepackage{xr-hyper}
\let\vec\mathbf   
\newcommand{\kin}{\vec{k}_\mathrm{in}}
\newcommand{\kout}{\vec{k}_\mathrm{out}}
\newcommand{\hkl}{{hk\ell}}
\DeclareSIUnit\angstrom{\text {Å}}
\DeclareSIUnit\photon{\text {ph}}

\begin{document}

\title[Article Title]{Convergent-Beam X-ray Crystallography} 

\author*[1]{\fnm{Chufeng} \sur{Li}}\email{chufeng.li@desy.de}
\author[1]{\fnm{Margarita} \sur{Zakharova}}\email{margarita.zakharova@desy.de}
\author[1]{\fnm{Mauro} \sur{Prasciolu}}\email{mauro.prasciolu@desy.de}
\author[1]{\fnm{Jia Chyi} \sur{Wong}}\email{wong.jia.chyi@desy.de}
\author[1]{\fnm{Holger} \sur{Fleckenstein}}\email{holger.fleckenstein@desy.de}
\author[2]{\fnm{Nikolay} \sur{Ivanov}}\email{nikolay.ivanov@desy.de}
\author[1]{\fnm{Wenhui} \sur{Zhang}}\email{wenhui.zhang@desy.de}
\author[2]{\fnm{Mansi} \sur{Butola}}\email{mansi.butola@desy.de}
\author[2]{\fnm{J. Lukas} \sur{Dresselhaus}}\email{lukasdresselhaus@web.de}
\author[1]{\fnm{Ivan} \sur{De Gennaro Aquino}}\email{ivan.degennaroaquino@gmail.com}
\author[1]{\fnm{Dmitry} \sur{Egorov}}\email{dmitry.egorov@maxiv.lu.se}
\author[1]{\fnm{Philipp} \sur{Middendorf}}\email{philipp.middendorf@desy.de}
\author[1]{\fnm{Alessa} \sur{Henkel}}\email{alessa.henkel@desy.de}
\author[2]{\fnm{Bjarne} \sur{Klopprogge}}\email{bjarne.klopprogge@desy.de}
\author[2,3]{\fnm{Lars} \sur{Klemeyer}}\email{lars.klemeyer@uni-hamburg.de}
\author[2,3]{\fnm{Tobias} \sur{Beck}}\email{tobias.beck@uni-hamburg.de}
\author[1]{\fnm{Oleksandr} \sur{Yefanov}}\email{oleksandr.yefanov@desy.de}
\author[1]{\fnm{Miriam} \sur{Barthelmess}}\email{miriam.barthelmess@desy.de}
\author[1,2]{\fnm{Janina} \sur{Sprenger}}\email{janina.sprenger@cfel.de}
\author[1,2]{\fnm{Dominik} \sur{Oberthuer}}\email{dominik.oberthuer@desy.de}
\author[1,2]{\fnm{Sa\v{s}a} \sur{Bajt}}\email{sasa.bajt@desy.de}
\author*[1,2,4]{\fnm{Henry N.} \sur{Chapman}}\email{henry.chapman@desy.de}

\affil[1]{\orgdiv{Center for Free-Electron Laser Science CFEL}, \orgname{Deutsches Elektronen-Synchrotron DESY}, \orgaddress{\street{Notkestr.\ 85}, \postcode{22607} \state{Hamburg}, \country{Germany}}}

\affil[2]{\orgname{The Hamburg Centre for Ultrafast Imaging}, \orgaddress{\street{Luruper Chaussee 149}, \postcode{22761} \state{Hamburg}, \country{Germany}}}

\affil[3]{\orgdiv{Department of Chemistry}, \orgname{University of Hamburg}, \orgaddress{\street{Grindelallee 117}, \postcode{20146} \state{Hamburg}, \country{Germany}}}

\affil[4]{\orgdiv{Department of Physics}, \orgname{University of Hamburg}, \orgaddress{\street{Luruper Chaussee 149}, \postcode{22761} \state{Hamburg}, \country{Germany}}}

\abstract{
Molecular and polymeric crystals show a wide range of functional properties that arise from the interplay between the atomic-scale structure of their constituent molecules and the organization of these molecules within the crystal lattice at macroscopic length scales. X-ray diffraction can provide structural information at these disparate length scales, but usually only through experiments that address one or the other of molecular (or unit-cell) structure versus crystal structure. Consequently, the accuracy of determined molecular or polymer structures may be limited by unaccounted crystal inhomogeneities of the crystal lattice and the characterization of crystalline materials might not reveal the underlying causes of crystal morphology. Here we introduce X-ray convergent-beam diffraction to obtain spatially-resolved structural information from crystals by projection topographic imaging. Using highly focusing X-ray multilayer Laue lenses, we show that Bragg reflections can be mapped into tomographic images of the crystal, for the characterization of strain and defects at high resolution. We demonstrate how the crystal morphology obtained this way can be accounted for when determining structure factors as a function of position in the crystal. The approach may assist in studies such as diffusion and binding in MOFS, protein-drug binding, crystal growth, and the mechanical responses of photo-reactive or thermally driven dynamic crystals. 

}
\keywords{Molecular crystallography; Topography; Diffraction contrast; Multilayer Laue lenses}

\maketitle

\section{Introduction}
\label{sec:intro}
Crystals formed from molecules, ranging from small compounds to macromolecules, or polymers and metal organic frameworks, are typically characterized by their crystal parameters of unit-cell dimensions and space group. This characterization, and the perfection to which a crystal conforms to it, enables a crystallographic diffraction analysis to measure structure factors and obtain an atomic model of the unit cell or molecular components of crystal. In reality, such crystals are never perfect, and variations in lattice parameters, such as strains and twists of the lattice, or defects and vacancies, modify the diffraction pattern in a variety of ways. Properly accounting for the effects of these inhomogeneities on measured structure factors requires knowing their cause, or at least how they vary or are distributed throughout the volume of the crystal. This is especially relevant for the characterization of molecular and polymeric crystals with dynamic properties that respond to electric, magnetic, and light fields, or external mechanical or thermal forces, and which may find use as switches and actuators, for flexible electronics, or novel photonic devices~\cite{Awad:2023}. Likewise, variations in the conformations and arrangements of molecules throughout a crystal, such as due to the diffusion of binding compounds or the absorption of an optical trigger pulse, may confound conventional analyses and instead require a spatially-resolved crystal diffraction analysis.

For crystals considerably thinner than \SI{100}{\nm}, convergent-beam electron diffraction (CBED) is well established in the scanning transmission electron microscope (STEM)~\cite{Spence:1992}, where the beam size can be easily controlled and reduced down to below the size of crystal unit cells or even single atoms. The approach is extensively used to measure strain, lattice order, grain boundaries, and map grain orientations or crystal ordering in hard materials such as minerals, nanomaterials, and devices~\cite{Savitzky:2021}.
X-rays are much more penetrating than electrons, enabling the analysis of macroscopic crystals. Position-resolved
X-ray diffraction has been applied by scanning micrometer-sized beams across crystals for mapping grains in polycrystalline materials. Approaches such as tensor tomography \cite{Korsunsky:2006,Liebi:2015,Carlsen:2024} and multigrain mapping~\cite{Poulsen:2001} combine measurements from many Bragg reflections to determine lattice orientations, but such approaches have not been used to determine molecular structure changes. Other approaches such as crystal topography~\cite{Topography-book,Fourme:1995}, diffraction microscopy~\cite{Simons:2015,Yildirim:2023},  Bragg coherent diffractive imaging~\cite{Pfeifer:2006a,Sun:2024} and Bragg ptychography~\cite{Hruszkewycz:2017} usually obtain spatially-resolved measurements at a single Bragg peak at a time.  
It was only when optics 
of high convergence became available that convergent-beam X-ray diffraction (CBXD) of molecular and macromolecular crystals was first demonstrated~\cite{Ho:1998,MacDonald:1999,Ho:2002}. However, those optics were far from the quality required for microscopy or illuminating small crystals, and the approach was not further developed. 

The advent of efficient multilayer Laue lenses (MLLs) of high numerical aperture~\cite{Yan:2014,Bajt:2018} brings new opportunities, closer to the situation of CBED in the STEM, which we explore here.
These lenses are sliced from a multilayer parent structure formed by physical vapor deposition to create a volume diffractive element with periods as small as \SI{1}{\nm}. 
Nanometer multilayer periods confer the ability to deflect hard X-ray beams by several degrees. For example, a lens used in this study has a semi-angle of convergence, $\alpha = \SI{1.6}{\degree}$ (or numerical aperture, $\mathrm{NA}=\sin\alpha = 0.028$) at a wavelength of \SI{0.071}{\nm}. This would result in a diffraction-limited resolution of about \SI{1.3}{\nm} if fabricated without imperfections (which cause lens aberrations~\cite{Dresselhaus:2024}). This can be comparable to the unit-cell dimensions of molecular crystals, implying that the lens convergence angle can exceed the angular separation of Bragg peaks. The convergence angle of MLLs is significantly larger than the rocking-curved width of Bragg peaks, producing ``Bragg streaks'' on the detector with a length dependent on $\alpha$. These are related to Kossel lines formed by the diffraction of fluorescence emitting in all directions and which encode structural information from a large volume of reciprocal space~\cite{Lonsdale:1947,Glazer:2015,Lider:2011}.

Just as in the electron microscope, CBXD using MLLs provides many modalities to extract structural information of crystalline samples. For example, by scanning the focused probe across the face of a crystal, the strength and position of Bragg streaks could be used to map the local diffraction efficiency and lattice orientation~\cite{Ophus:2019}. By combining this with rotation of the crystal, complete sets of structure factors could be obtained from different sub-volumes of the crystal. In practice, using synchrotron radiation, the high intensity of the focused probe can quickly damage radiation sensitive materials and we find that spatially-resolved diffraction data can be collected more easily by placing the crystal in the defocused beam, which is much wider than the focused beam and hence not as intense. By placing the crystal in the expanding beam, the transmitted and diffracted beams project magnified images onto a detector placed further downstream.  We show how magnified diffraction topographs can be obtained for an almost complete set of reflections and that these can be used to obtain a tomographic image of the crystal. We detail how the photon counts in these topographs can be integrated to obtain accurate structure factors, and demonstrate the possibility of determining molecular structures as a function of position within the crystal.   

\section{Convergent-beam diffraction}
\begin{figure}[!tb]
    \centering
    \includegraphics[width=0.95\linewidth]{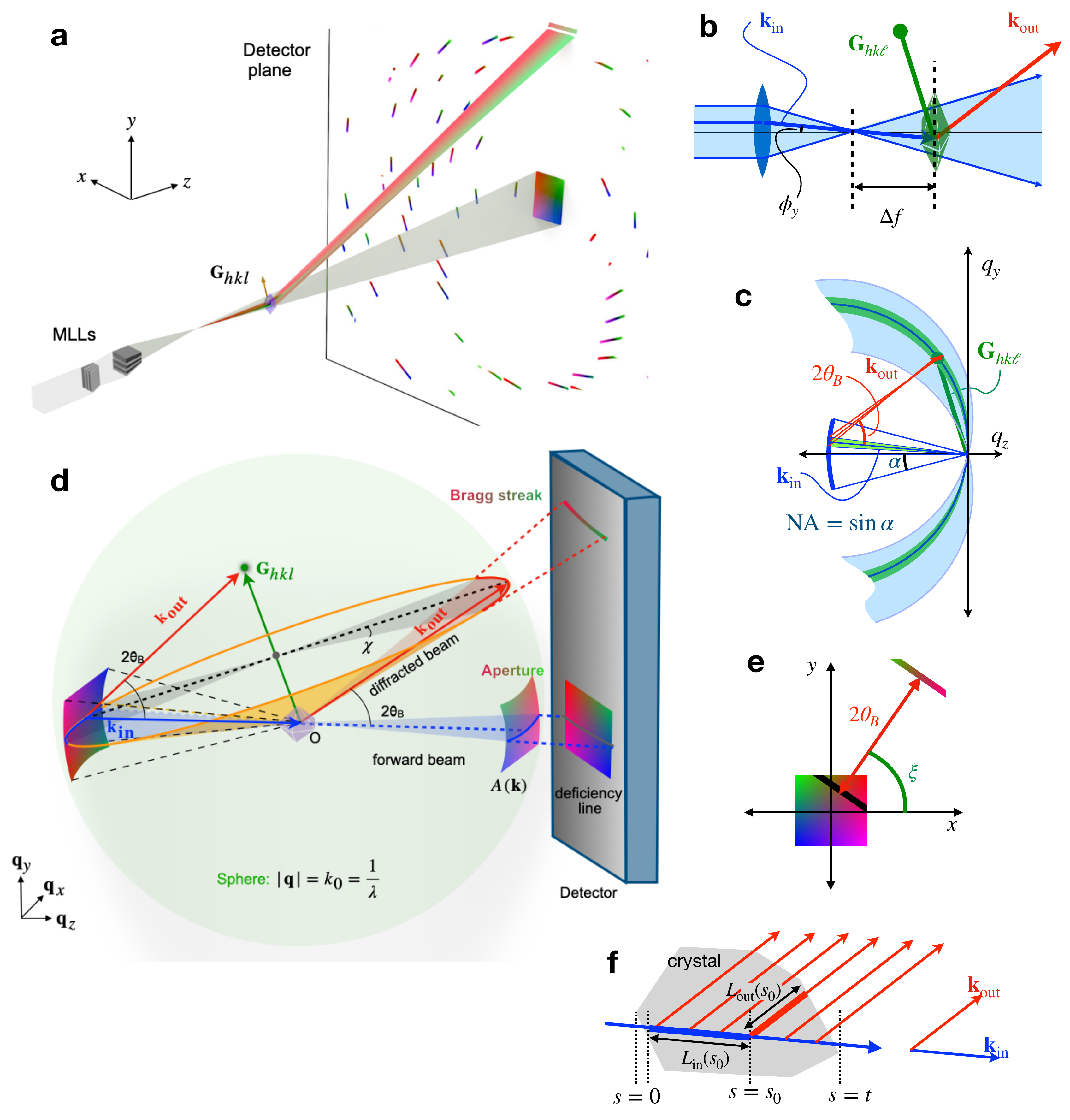}
    \caption{The geometry of CBXD. (a) A crystal is placed in the beam focused by an X-ray lens (here, a pair of MLLs) and the diffraction pattern is recorded on a plane pixel-array detector. (b) Of all rays provided by the lens, only those that satisfy the diffracting condition for a RLP $\vec{G}_{\hkl}$ will reflect. A ray originating from an angle $\phi_y$ to the optical axis maps to a position $\Delta f\tan \phi_y$ on a crystal defocused by $\Delta f$. (c) A volume in reciprocal space (blue shading) is formed by Ewald spheres rotated over a range of $\kin$ wave-vectors as supplied by the lens. Any $\vec{G}_{\hkl}$ within this volume will produce a reflection, $\kout$. (d) The diffraction condition for $\vec{G}_{\hkl}$ is satisfied for all wave-vectors that lie on a Kossel circle $\vec{k}_K(\chi)$ (shown in orange) in a plane perpendicular to and bisecting $\vec{G}_{\hkl}$. A Bragg streak and a deficiency line are generated if $\vec{k}_K(\chi)$ passes through the lens aperture. (e) The deficiency line (in the projected lens pupil) and Bragg streak are approximately parallel on the detector and separated by an angle $2\theta_B$. (f) Crystal diffraction for a given incident ray path is found by integrating over all paths $L_\mathrm{in} + L_\mathrm{out}$ for all possible scattering locations $s_0$ along the incident path, assuming single scattering. } 
    \label{fig:CBXD_geometry}
\end{figure}

Our CBXD scheme is depicted in Fig.\ \ref{fig:CBXD_geometry} (a), where diffraction patterns are recorded directly on a pixel array detector as the sample is scanned and rotated. The instrument is in all respects a scanning transmission X-ray microscope (STXM)~\cite{Jacobsen-book}, although here we place the sample out of focus. In this case, the beam transmitted through the crystal forms a magnified projection image on the detector with a magnification given by the ratio of the focus to detector distance, $l_D$, to the crystal defocus distance, $\Delta f$~\cite{Zhang:2024}. As seen in Figs.\ \ref{fig:CBXD_geometry} (a) and (b), there is a similar magnification encoded along the length of the Bragg streaks, which we show can be used to construct projection diffraction topographs---maps of the diffraction efficiency---either from all streaks in a single static diffraction pattern or individually from each Bragg streak as the crystal is rotated such that the diffracting condition sweeps across the face of the crystal. 

\subsection{Geometric model for convergent-beam diffraction}
\label{sec:CBD}
In the kinematical formalism, diffraction of a collimated monochromatic incident beam of wave-vector $\kin$ from the periodic structure of a crystal  occurs only in those directions $\kout=\kin+\vec{G}_\hkl$ where the reciprocal lattice points (RLPs) $\vec{G}_{\hkl}$ intersect the Ewald sphere---a manifold of radius $1/\lambda$, centered on $-\kin$ and thus intersecting the reciprocal-space origin~\cite{Cowley:1981,Als-Nielsen}. (Here, $|\kin| =1/\lambda$ for a wavelength $\lambda$.) A lens supplies a range of incident ray directions $\kin$, each giving rise to a tilted Ewald sphere. Together, these form a volume of reciprocal space depicted as the blue shaded region in Fig.~\ref{fig:CBXD_geometry}~(c). All RLPs in that volume contribute to the diffraction pattern. This situation is akin to Laue diffraction, where a polychromatic beam fills out a reciprocal space volume limited by Ewald spheres of different radii dependent on the minimum and maximum wavelengths. 
It was shown that for scattering angles $2\theta$ up to \SI{45}{\degree}, the reciprocal-space volume contributing to a CBXD pattern formed with a lens of $\mathrm{NA} = 0.03$ is equivalent to Laue diffraction with a bandwidth of \SI{33}{\percent}~\cite{Chapman:2025}, bringing a corresponding increase in the number of reflections in diffraction patterns of static crystals. 
The large excitation volume of CBXD makes it especially attractive for serial crystallography of small molecules where snapshot patterns of small-unit-cell crystals have proved difficult to index due to a sparsity of Bragg spots~\cite{Moon:2024}.    

A significant difference between polychromatic and convergent-beam diffraction is that in Laue diffraction, wavelength is a scalar quantity, whereas the angles $(\phi_x,\phi_y)$ provided by a lens are two dimensional. This extra degree of freedom produces Bragg streaks in CBXD patterns instead of spots in polychromatic diffraction patterns. Essentially, if a ray $\kin$ supplied by the lens satisfies the Bragg diffraction condition for a particular $\vec{G}_{\hkl}$, then rays formed by rotating $\kin$ around the axis defined by $\vec{G}_{\hkl}$ will all reflect with the same Bragg angle $\theta_B$ to the corresponding lattice planes. These rays lie on a cone, known as the Kossel cone~\cite{Frank:1972}, shown in Fig.~\ref{fig:CBXD_geometry} (d), with a semi-angle $\pi/2-\theta_B$ and an axis that coincides with $\vec{G}_{\hkl}$ 
We describe the cone by vectors $\vec{k}_K(\chi)$ which trace out the orange circle in Fig.~\ref{fig:CBXD_geometry}~(d) as a function of the azimuthal angle $\chi$. 

We follow the analysis formalism of Kossel patterns~\cite{Frank:1972} to determine an expression for the ``Kossel circle'', $\vec{k}_K(\chi)$. Since the magnitudes of $\kin$ and $\kout$ are equal, the average vector $\bar{\vec{k}} = (\kin+\kout)/2$ bisects and is perpendicular to the difference $\kout-\kin$. Thus, the full circle that contains the Bragg streak $\kout(\chi)$ and the line of incident wave-vectors $-\kin(\chi)$ converging on the origin, for which $\kout-\kin=\vec{G}_{\hkl}$, is found as the intersection of the sphere of radius $1/\lambda$ centered at the origin of reciprocal space (the ``X-ray Fermi sphere''), with the plane that is normal to and bisects $\vec{G}_{\hkl}$ (the ``Brillouin plane''), giving rise to the orange circle seen in Fig.~\ref{fig:CBXD_geometry}~(d) that defines the Kossel cone~\cite{Frank:1972}. The Kossel circle is therefore defined by coordinates $\vec{q}$ that simultaneously satisfy the two equations~\cite{Chapman:2025} 
\begin{equation}
    \vec{q}^2 = \frac{1}{\lambda^2}; \quad \left( \vec{q} - \frac{\vec{G}_{\hkl}}{2}\right)\cdot \vec{G}_{\hkl} = 0,
    \label{eq:circle}
\end{equation}
and which can be parameterized by the angle $\chi$ as
\begin{equation}
    \label{eq:kp}
    \vec{k}_K(\chi) = \frac{\vec{G}_{\hkl}}{2}+\cos \chi \,\vec{a}+\sin \chi \, \vec{b},
\end{equation}
where the radius of the circle is given by $|\vec{a}| = |\vec{b}| =|\bar{\vec{k}}|  = (1/\lambda)\cos \theta_B$ for a Bragg angle $\theta_B$. The zero of the angle $\chi$ is arbitrary, but set here by taking the vector $\vec{a}$ to be in the plane containing both $\vec{G}_{\hkl}$ and the optical axis, $\hat{\vec{z}}$. That is, $\vec{a} \cdot (\vec{G}_{\hkl} \times \hat{\vec{z}})=0$, for which $a_x\, G_y = a_y\, G_x$ for the vector components of $\vec{a}$ and $\vec{G}_{\hkl}$ in the laboratory frame. Since $\bar{\vec{k}}$ is orthogonal to $\vec{G}_{\hkl}$, then $\vec{a} \cdot \vec{G}_{\hkl} = 0$, which together give the identity $(a_x^2+a_y^2)/a_z^2 = G_z^2/(G_x^2+G_y^2)$ and from which it can be found that  
\begin{equation}
\label{eq:a}
\begin{split}
     a_z^2&=\frac{1/\lambda^2 - \vec{G}^2_{\hkl}/4}{1+G_z^2/(G_x^2+G_y^2)} = \frac{1}{4}(G_x^2+G_y^2)\cot^2 \theta_B \\
    a_x &=  -a_z \frac{G_z\,G_x}{G_x^2+G_y^2} \\
    a_y &=  -a_z \frac{G_z\,G_y}{G_x^2+G_y^2}. 
\end{split}
\end{equation}
The vector $\vec{b}$ is orthogonal to both $\vec{a}$ and $\vec{G}_{\hkl}$, so its direction can be computed from the cross product $\vec{a} \times \vec{G}_{\hkl}$. Thus, $\vec{k}_K(\chi)$ is fully determined from $\vec{G}_{\hkl}$ for a given wavelength, as computed using Eqns~(\ref{eq:kp}) and (\ref{eq:a}).

The range of angles $\chi$ that define the Bragg streak $\kout(\chi) = \vec{k}_K(\chi)$ is limited by the intersection of incident $\kin(\chi) = -\vec{k}_K(\pi-\chi)$ with the lens aperture, displayed in Fig.~\ref{fig:CBXD_geometry} (d), and which reflect to the Bragg streak. With sufficiently thick crystals, extinction of the incident beam due to diffraction causes a dark line in projected beam on the detector, giving the name ``deficiency line''~\cite{Uesugi:2024,Morniroli:2004}. Extinction is usually not detectable for the small molecular crystals examined here.  Analysis of a CBXD pattern therefore requires determining from the Bragg streaks both the reciprocal lattice index and the angular coordinates of the deficiency line $\kin(\chi)$ (constrained by the exit pupil of the lens), analogous to determining the index and wavelength for each spot in a Laue diffraction pattern (constrained by the spectrum). 
The deficiency line and Bragg streak are approximately parallel at the detector, as depicted in Fig.~\ref{fig:CBXD_geometry}~(e), and the angular separation of each is always $2\theta_B$.

The short width of the Bragg streak gives the rocking curve of the reflection. Modeling the crystal as a slab of thickness $t$ produces a reciprocal lattice profile along the $q_z$ direction that is proportional to $\mathrm{sinc}^2(\pi q_z t)$ with a half width of $1/t$, independent of the distance of the RLP to the origin. As seen from Fig.~\ref{fig:CBXD_geometry} (c), there is a range of $\kin$ wave-vectors, associated with a range of Ewald spheres (green shaded region), that intersect the reciprocal lattice profile. This angular range is
\begin{equation}
    \label{eq:rocking}
    \Delta \theta = \frac{\lambda}{t \sin 2 \theta_B} =\frac{\lambda}{t}L, 
\end{equation}
where $L=1/\sin2\theta_B$ is the Lorentz factor, which indicates how the angular range of Bragg streaks varies across the pattern. For a snapshot pattern, the proportion of the incident flux supplied by the lens that is directed into a particular Bragg streak is proportional to $\Delta \theta$ and hence to $L$. 
We note that for the lens NA and crystals considered here, the streak lengths (along $\chi$) are about 100 times their widths, $\Delta \theta$. The short width is thus unlikely to be cut by the lens aperture, giving total counts in the streaks that are proportional to fully integrated intensities~\cite{Chapman:2025}.   

\begin{figure}[!tb]
    \centering
    \includegraphics[width=0.75\linewidth]{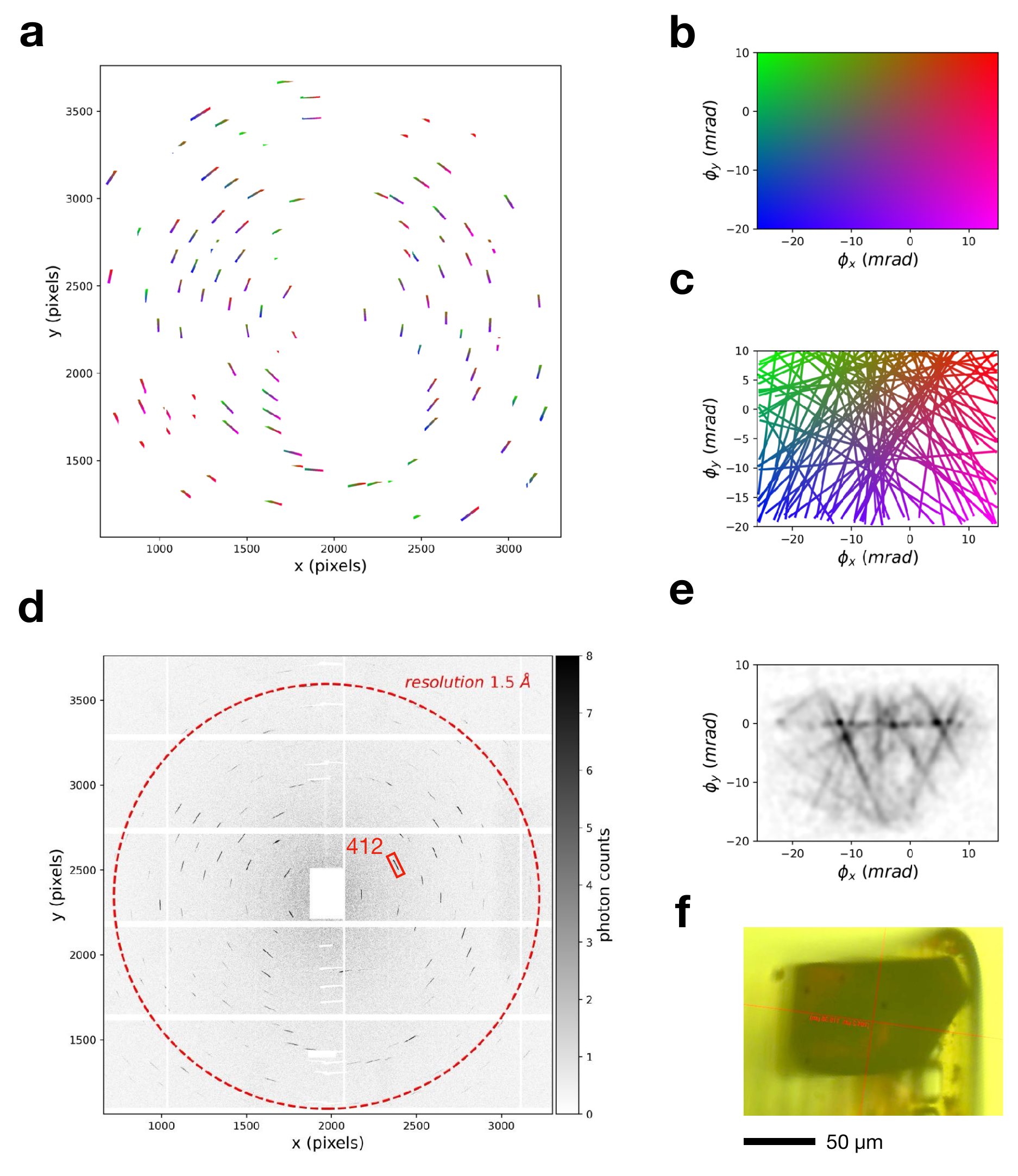}
    \caption{Vitamin B$_{12}$ snapshot convergent-beam diffraction. (a) A calculated diffraction pattern for a photon energy of \SI{17.5}{\kilo\eV}, with Bragg streaks coloured according to the position of the $\kin$ wave-vectors in the lens aperture as mapped in (b). The corresponding deficiency lines are mapped across the lens pupil in (c). (d) Experimental CBXD pattern of a vitamin B$_{12}$ crystal recorded with a lens of $\mathrm{NA} = 0.028$ at a defocus of $\Delta f =\SI{3.5}{\mm}$, and in the same orientation as calculated in (a). (e) The mapping of the measured Bragg streaks to their corresponding incident $\kin$ wave-vectors in the lens pupil. (f) Optical image of the crystal from the beamline in-line microscope for the crystal orientation used for (d).}
    \label{fig:B12-snapshot}
\end{figure}

\subsection{Snapshot diffraction pattern}
A computed snapshot (or still) pattern of Bragg streaks for the orthorhombic lattice of a stationary vitamin B$_{12}$ crystal is shown in Fig.~\ref{fig:B12-snapshot}~(a) for a lens aperture with convergence semi-angles $\alpha$ of \SI{0.021}{\radian} (horizontal) and \SI{0.015}{\radian} (vertical) and a photon energy of \SI{17.5}{\kilo\eV}. The corresponding mapping of the deficiency lines is given in Fig.~\ref{fig:B12-snapshot}~(c). For this illustration, the angular coordinates of $(\phi_x,\phi_y)$ of the lens aperture are encoded by a colour map displayed in Fig.~\ref{fig:B12-snapshot}~(b). The deficiency lines in Fig.~\ref{fig:B12-snapshot}~(c) map directly to this colour scheme and these colours are replicated along each Bragg streak in Fig.~\ref{fig:B12-snapshot}~(a). In this pattern, there are 96 Bragg streaks to a resolution of \SI{1.5}{\angstrom}. It is clear from this visualisation that if the lens aperture is stopped down, fewer (and shorter) Bragg streaks will be observed in the pattern. For example, if only rays were supplied from within the red corner of the aperture, only the red parts of the diffraction pattern would appear. A collimated beam may only excite one or two reflections, if any. An experimental CBXD pattern of a stationary vitamin B$_{12}$ crystal is shown Fig.~\ref{fig:B12-snapshot}~(d), for the same orientation and wavelength but a lens with even higher convergence of $\alpha = \SI{0.028}{\radian}$ in the horizontal and vertical directions   (see Methods, Sec.~\ref{sec:methods-instrument}). 

Despite the high lens convergence, some streaks of the experimental pattern of Fig.~\ref{fig:B12-snapshot} (d) are shorter than their counterparts in the calculated pattern. This is because the crystal was placed downstream of focus where the beam over-filled the crystal. In this case, the incident wave-vectors $\kin$ map to positions in the defocused beam, and hence to positions across the face of the crystal as depicted in Fig.~\ref{fig:CBXD_geometry} (b). Where a deficiency line extends beyond the boundary of the crystal, diffraction obviously does not occur and the corresponding Bragg streak will be truncated. Indeed, the intensity along the Bragg streak depends on the strength of diffraction experienced by rays of $\kin$. In the kinematical formalism, equivalent to the assumption of single scattering within the Born approximation, the integrated diffraction efficiency depends only on the crystal structure projected along the \emph{incident} ray path.

The intensity along the streak, $I_\hkl(\kout(\chi))$, can be determined using a simple model of the crystal as a 3D shape $S(\vec{x})$ which has the value of 1 in a volume element that has perfect crystallinity, and 0 where there is no crystal order. For a perfect crystal, the integral of $S(\vec{x})$ is equal to the volume of the crystal. As long as Bragg streaks do not  overlap in a CBXD pattern (and thus do not interfere), then the counts along the Bragg streak, integrated across the rocking direction, can be written as
\begin{equation}
    I_\hkl(\kout(\chi)) = c \,P\,L\,|F_\hkl|^2 \,|A(\kin(\chi))|^2\, {\cal{P}}_S (\kin(\chi),\omega)
\label{eq:intensity}    
\end{equation}
where $F_\hkl$ is the structure factor, $A$ the aperture function of the lens, $P$ the usual polarisation factor, $L$ the Lorentz factor of Eqn.~\ref{eq:rocking}, and $c$ is a constant. The orientation of the crystal is parameterised here by the rotation angle $\omega$ around a single fixed axis but any rotation matrix $\mathbf{R}$ can be applied to $S(\vec{x})$. The projection of the crystal shape is thus along all paths $p(\kin)$ along rays $\kin(\chi) = \kout(\chi) - \vec{G}_\hkl = -\vec{k}_K(\pi-\chi)$, and given by
\begin{equation}
    {\cal{P}}_S(\kin,\omega) = \int_{p(\kin)} \mathbf{R}_\omega S(\vec{x}) \,\mathrm{d}s. 
\label{eq:projection}
\end{equation}

\subsection{Analysis of snapshot CBXD patterns}
\label{sec:snapshot-analysis}
Equation~\ref{eq:intensity} shows the integrated intensities of the Bragg streaks carry information both of the structure factor square magnitudes $|F_\hkl|^2$ and of projections through the 3D crystal morphology $S(\vec{x})$. When the crystal is placed out of focus, paths $p(\kin(\chi))$ make a fan projecting from the focal point that slices through the crystal as depicted in Fig.~\ref{fig:CBXD_geometry} (b). This fan projects view on the detector, magnified by $l_D/\Delta f$.
A partial reconstruction of a two-dimensional magnified projection image ${\cal{P}}_S$ of the crystal can therefore be formed by mapping each Bragg streak back to its incident deficiency line $\kin(\chi)$, after first indexing the streaks in the diffraction pattern, as shown in Fig.~\ref{fig:B12-snapshot}~(e). The image is a rather sparse representation of the crystal, but the shape of the crystal is revealed, in agreement with the optical image of Fig.~\ref{fig:B12-snapshot} (f). The mapping is displayed in terms of the angular coordinates $(\phi_x,\phi_y)$ of the incident beam. Given the defocus distance $\Delta f=\SI{3.5}{\mm}$, each \SI{75}{\micro\meter} wide pixel in the detector maps to a spacing of about \SI{1.4}{\micro\meter} at the plane of the crystal. 

\begin{figure}[!tb]
    \centering
    \includegraphics[width=1.0\linewidth]{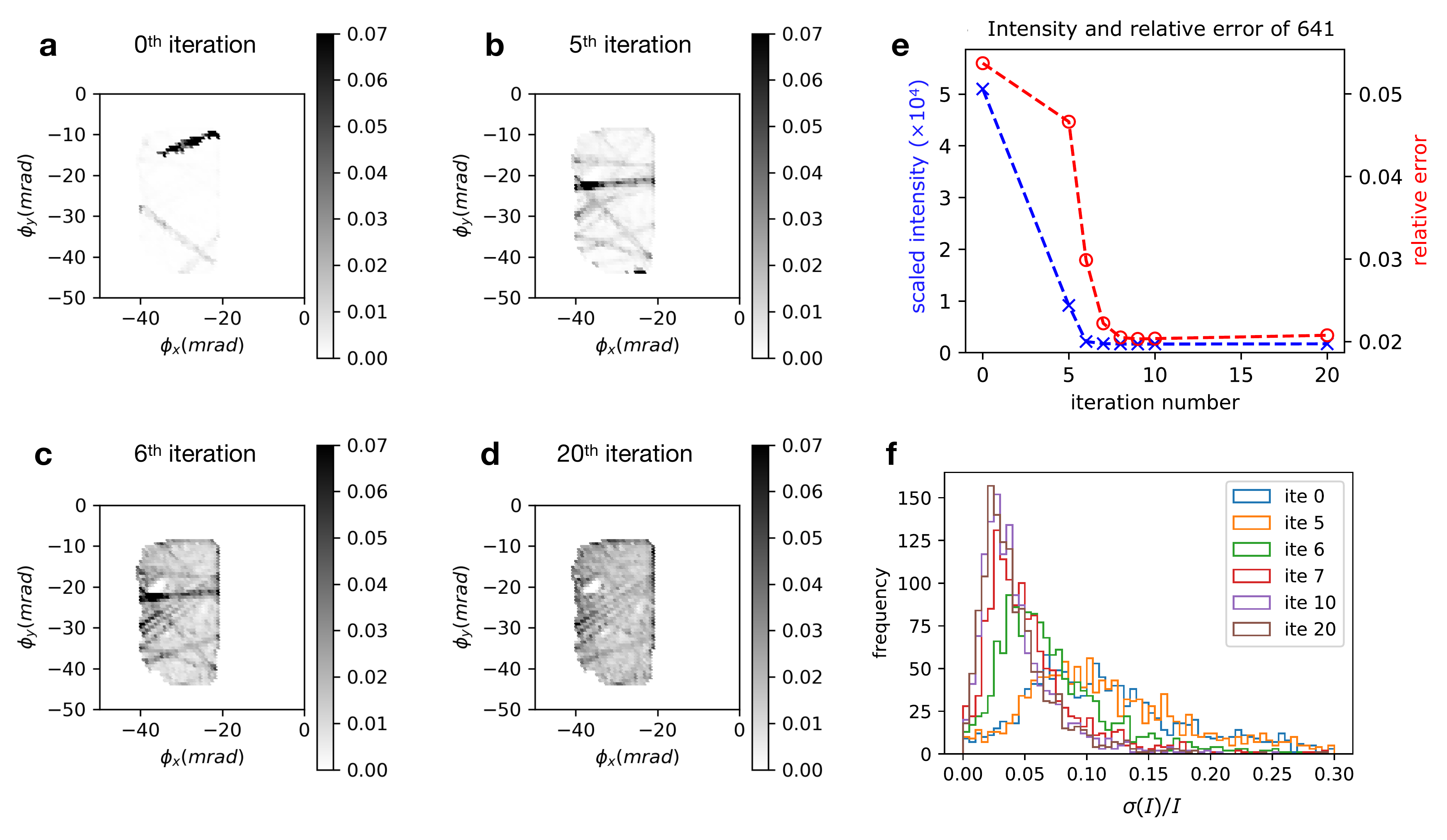}
    \caption{Estimates of a topograph of a vitamin B$_{12}$ crystal determined from a single snapshot CBXD pattern. (a) Mapping of measured Bragg streaks to the positions of their deficiency lines, before normalisation. (b)--(d) Iterates of the topograph, as consistency between overlapping deficiency lines, and reflection intensities of  same reflections across patterns is enforced. (e) Scaled intensity (blue) and relative error (red) of the 641 reflection, showing convergence after 8 iterations. (f) Histograms of the relative error of all observed reflections for various numbers of iterations. }
    \label{fig:bootstrap}
\end{figure}

The map of ${\cal{P}}_S(x,y)$ in Fig.~\ref{fig:B12-snapshot}~(e) was formed by normalising each measured Bragg streak by $|F_\hkl|^2$ and the aperture function $|A|^2$, as prescribed by Eqn.~(\ref{eq:intensity}). However, this first requires determining the structure factor magnitudes. These can be estimated from the relative strengths of the Bragg streaks, but only after knowing the spatial variation of the projected crystal diffraction strength ${\cal{P}}_S$. Self consistency can be achieved by adopting a simple iterative procedure that initially normalises each streak to a mean count of 1.0 per pixel which are then combined into a map by averaging values of ${\cal{P}}_S$ where they overlap. New estimates of ${\cal{P}}_S(\kin(\chi))$ are sliced out of that map for each streak and compared with the observations $I_\hkl(\kout(\chi))$ to update both $|F_\hkl|^2$ and ${\cal{P}}_S(\kin(\chi))$ by replacing ${\cal{P}}_S(\kin(\chi))$ with a rescaled version of $I_\hkl(\kout(\chi))$. To imagine how the intensities are constrained, consider three non-colinear streaks whose deficiency lines intersect at three points that form a triangle. An example of the process is illustrated in Fig.~\ref{fig:bootstrap} for another vitamin B$_{12}$ crystal. Here the crystal was shaped as a prism and was quite uniform in thickness. 

The final map of ${\cal{P}}_S(x,y)$ in Fig.~\ref{fig:bootstrap} (d) contains some holes in locations where no deficiency lines crossed and so where no information is provided by the snapshot CBXD pattern. The iterations are monitored by convergence of the scaling of each $I_\hkl(\kout(\chi))$ to provide the estimates of Bragg intensities, $\bar{I}_\hkl$, as shown in Fig.~\ref{fig:bootstrap} (e) for the 641 reflection. The relative error of the estimated signal of each reflection, $\sigma(\bar{I})/\bar{I}$, is also plotted. This error is determined from the standard deviation of the counts measured in pixels contributing to the Bragg streak after normalising by the updated ${\cal{P}}_S(\kin(\chi))$. As seen in the histogram of the relative errors of all observed reflections in the diffraction pattern, shown in Fig.~\ref{fig:bootstrap} (f), achieving consistency significantly reduces the relative error of all reflections.

\section{Rotation CBXD}
\label{sec:rotation-CBXD}
The analysis of Sec.~\ref{sec:snapshot-analysis} will be of use for serial snapshot crystallography, where only single patterns are obtained for each crystal, and especially if patterns are recorded with the defocused beam. 
However, in that case the projection map of the crystal is a composite from a large number of reflections, from which quantitative information about strain or lattice defects may be difficult to quantify. We show here that projection topographs can be acquired in parallel from many reflections in a rotation scan of the crystal, and that structure factors can then be estimated by integrating counts in those topographs or from selected regions of the crystal.

\subsection{Rotation topographs}
\label{sec:topo-rotation}
The deficiency line cuts across the face of a crystal placed out of focus and so a topograph of the crystal for a particular reflection can be formed either by moving the crystal transversely, so that the deficiency line scans across the entire crystal face, or by rocking the crystal about a single axis, which sweeps the deficiency line in a similar way across the crystal face.  The advantage of rocking the crystal is that the entire volume of the crystal can remain fully illuminated by the defocused beam throughout the entire scan and thus all volume elements of the crystal contribute to the dataset at every crystal orientation.  

\begin{figure}[!tb]
    \centering
    \includegraphics[width=\linewidth]{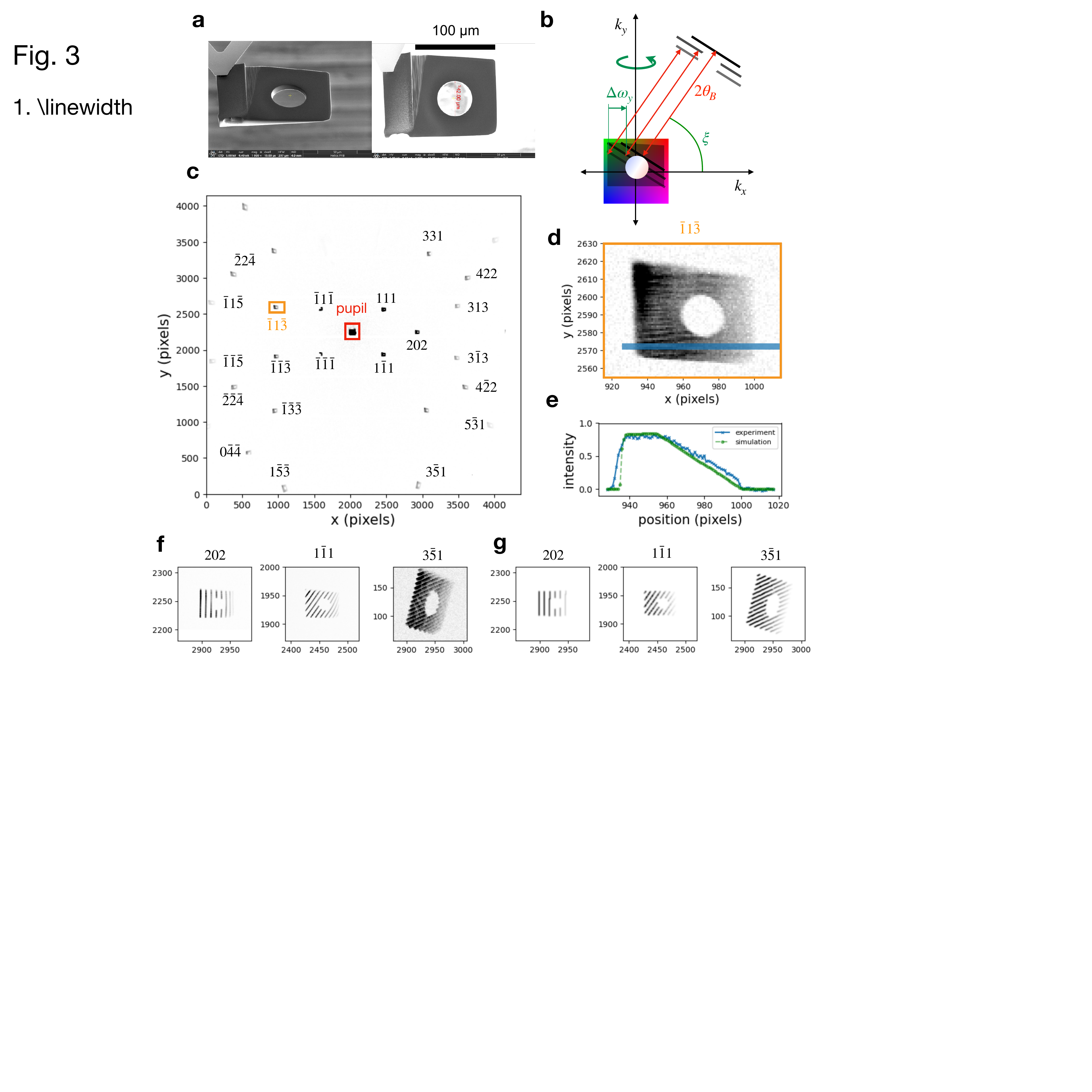}
    \caption{Rotation CBXD topographs of a silicon wedged single crystal with a hole, cut as a perpendicular lamella from a (100) Si wafer using a focused ion beam (FIB). (a) Scanning electron micrographs of the crystal.  (b) The deficiency line and Bragg streak remain separated by $2\theta$ and move together (here horizontally) as the crystal is rotated, to map out the projected view of a defocused crystal. (c) Rotation diffraction pattern formed by summing exposures over a range of $\pm\SI{15}{\degree}$ about the vertical axis with a step size of \SI{0.02}{\degree}. (d) A distorted topogram from the $\bar{1}1\bar{3}$ reflection. (e) Line profiles of the measured and simulated topogram intensity along the positions indicated in (d). Experimental (f) and simulated (g) rotation topographs obtained with a coarse rotation step of \SI{0.2}{\degree}.}
    \label{fig:si_wedge_hole_topo}
\end{figure}

Figure~\ref{fig:si_wedge_hole_topo} illustrates the formation of rotation topographs for a small wedged Si crystal with a circular hole, chosen to give a low number of reflections and placed \SI{4.5}{\milli\meter} downstream of the focus of an off-axis MLLs with 0.015 NA. Here, the crystal was rotated about the vertical axis, corresponding to the (010) lattice direction.
When the crystal is rotated, the entire Kossel circle is also rotated. A Bragg streak will only be visible as long as the Kossel circle intersects the lens aperture, which will occur for a range of orientations as depicted in Fig.~\ref{fig:si_wedge_hole_topo} (b). 
As the rotation of the crystal is continued, other RLPs will rotate into and through the participating volume of reciprocal space. Each Bragg streak can thus be tracked and aggregated into its own magnified topograph of the crystal, as shown in Fig.~\ref{fig:si_wedge_hole_topo} (c) for a scan over a range of \SI{30}{\degree} with a step size of \SI{0.02}{\degree}, for a detector distance of \SI{20}{\cm}.  
Over this scan, more than 25 reflections were observed to a resolution of \SI{0.9}{\per\angstrom}.     

Zooming in to any of these reflections reveals an unmistakable image of the perforated crystal that is magnified by about 45 times as shown in Fig.~\ref{fig:si_wedge_hole_topo} (d) for the $\bar{1}1\bar{3}$ reflection. It also indicates a variation of diffraction efficiency across the crystal due to its wedged thickness, in agreement with a simulated rotation pattern based on a model of the crystal obtained from the SEM images and assuming kinematical diffraction. Line-outs of intensities along a horizontal direction from the measured and simulated topographs are compared in Fig.~\ref{fig:si_wedge_hole_topo}~(e). 

Topographs built from a much coarser step size of \SI{0.2}{\degree} are shown in Fig.~\ref{fig:si_wedge_hole_topo}~(f) and compared with simulations in Fig.~\ref{fig:si_wedge_hole_topo}~(g). The density of Bragg streaks varies, even though the rotation step is the same for all. This is because they step in the horizontal direction as the crystal rotates about the vertical ($y$) axis. As apparent in Fig.~\ref{fig:si_wedge_hole_topo} (b), the density of streaks depends on the in-plane orientation $\xi$ of the deficiency line relative to the rotation axis since their separation in the direction perpendicular to the lines is proportional to $W = 1/\cos\xi$. The relative intensities of the topographs are thus proportional to this ``sampling factor'', which is similar to a Lorentz velocity factor in rotational crystallography 
When $\xi = \pi/2$, then $\vec{G}_\hkl$ is oriented along the rotation axis and stays within the reflecting condition for an entire rotation.

An obvious feature of the topograms in Fig.~\ref{fig:si_wedge_hole_topo} is that they are geometrically distorted by being recorded at high diffraction angles on a flat detector. For example, the $\bar{1}1\bar{3}$ topograph of the Si crystal in Fig.~\ref{fig:si_wedge_hole_topo} (d) appears stretched along the diagonal. These distortions can be corrected simply by mapping each Bragg streak of the composite topogram back to its corresponding deficiency line, to match the projected hologram of the crystal recorded at low angles.
The topograph can therefore be written as
\begin{equation}
    I_\hkl(\kin(\phi_x,\phi_y)) = c \,P\,L\,W\,|F_\hkl|^2 \,|A(\phi_x,\phi_y)|^2\,{\cal{P}}_S(\phi_x,\phi_y;\omega)
    \label{eq:2D-topo}
\end{equation}
where $(\phi_x, \phi_y)$ are the transverse angular coordinates of the aperture function (and of the rotation topograph).

Note that although rotation topographs from neighbouring reflections may overlap on the detector, as long as Bragg streaks do not overlap in a still exposure they can still be separated. 
We must qualify, however, that the topograms are not strictly magnified projections of the crystal as illuminated by rays originating from a single source point. This is true for rays spreading along the length of a Bragg streak in a diffraction pattern of a stationary crystal, where $\kin(\chi)$ makes a fan that projects onto the detector, but as the crystal is rotated the Bragg condition is maintained through the sweep, so the $\kin$ and $\kout$ vectors remain at a fixed orientation relative to the crystal planes. Thus, in the frame of reference of the crystal, the topograph is composed of separate fan-planes that are all parallel to each other, which we refer to as a ``stacked projection''. In the laboratory frame the source point of the projection either rotates about the rotation axis as the crystal is rocked or is fixed in space as the crystal is translated.

\subsection{Topographs of strained crystals}
\begin{figure}[!tb]
    \centering
    \includegraphics[width=0.75\linewidth]{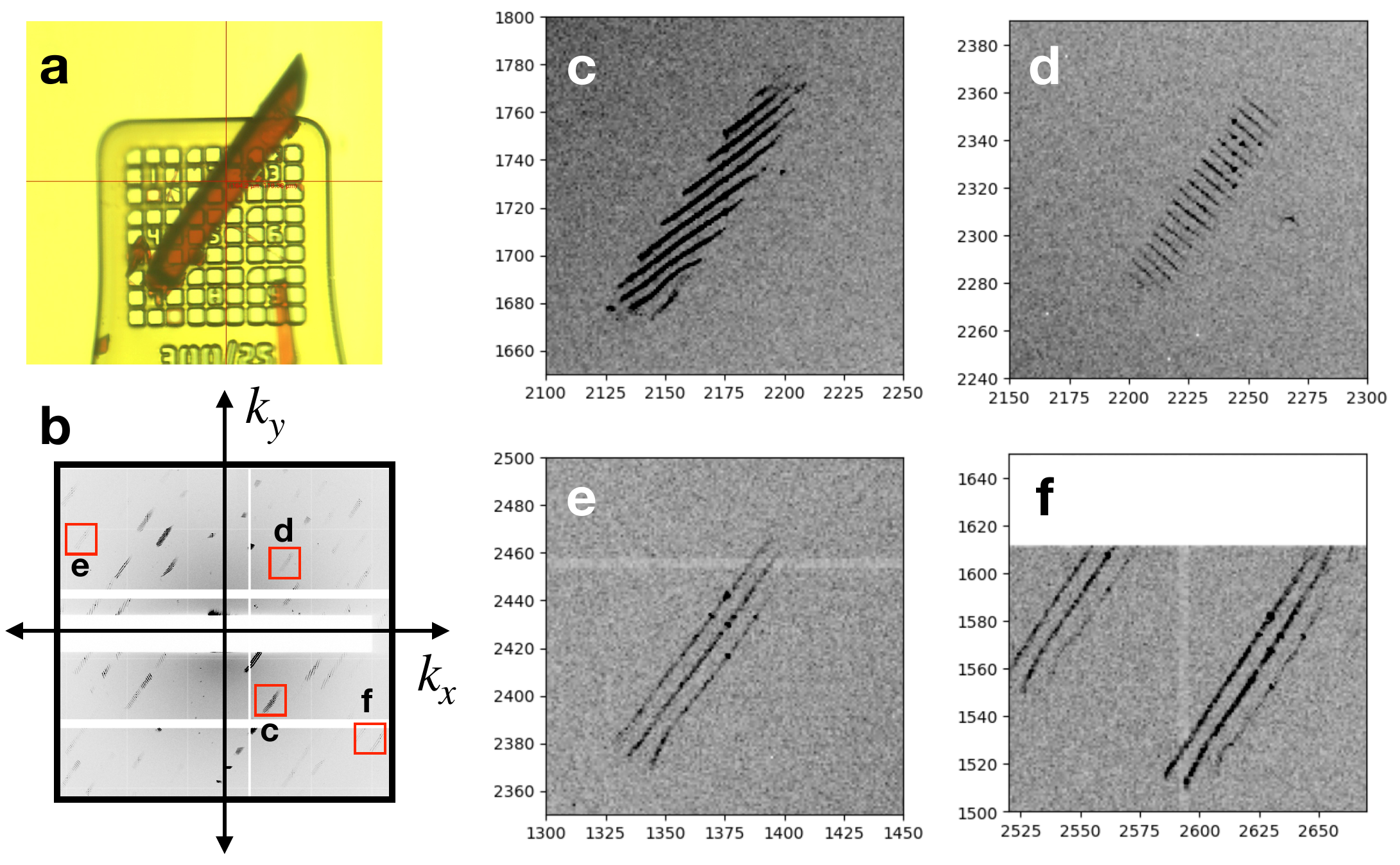}
    \caption{Coarse-step rotation topographs recorded with \SI{0.2}{\degree} step size. (a) In-situ micrograph of a needle-like vitamin B$_{12}$ crystal near the orientation used to collect a series of diffraction patterns (b), in which the positions of four topographs are indicated, shown in detail in (c) to (f). The detector pixel coordinates of the topographs are displayed. }
    \label{fig:B12-strain}
\end{figure}

Topographs are often used for mapping out lattice defects in crystals due to contrast changes these cause in the reflected intensities~\cite{Topography-book}. While the examples considered in this paper are from almost perfect crystals, we can consider the case of a strain field that varies along the length of the deficiency line cutting through the crystal. In this case, since the lattice spacing or orientation changes with position along the streak, the diffraction angle will also change, giving rise to a bend of the Bragg streak---an exemplar of mapping lattice changes in real space, and similar to an approach in CBED~\cite{Uesugi:2024}. An example is shown in Fig.~\ref{fig:B12-strain}, showing a coarsely-sampled rotation topograph of a long and thin vitamin B$_{12}$ crystal. The Bragg streaks near the bottom of the crystal are bent, indicating either a change in the unit-cell dimensions or a rotation of the lattice, or both. A comparison of the topographs (e) and (f), due to reflections in nearly opposite directions, suggest the distortion is due to a twist of the lattice.  
The coarse sampling of the topographs was made here just to illustrate how strain is encoded in the measurements. Higher-sampled strain fields can be constructed from a measure of the deviation of each Bragg streak from a straight line. 

\subsection{Tomo-topographs}
\label{sec:tomo-topo}
The topograms of the different reflections of the wedged silicon crystal in Fig.~\ref{fig:si_wedge_hole_topo} (c) occur for different crystal orientations as the corresponding RLP passes through the participating reciprocal space volume of the convergent beam. Unlike the snapshot topograph of Fig.~\ref{fig:B12-snapshot} (e), they present different projected views through the crystal. 
In general, however, strain and defects in the crystal will produce a contrast in the topograph that depends on the orientation relative to the incident beam, requiring the methodology of tensor tomography~\cite{Korsunsky:2006,Carlsen:2024} to obtain a complete 3D description. 


Here, instead, we solve the simpler problem of reconstructing the volume $S(\vec{x})$ of a perfect crystal that can be described by Eqn.~(\ref{eq:intensity}). That is, every reflection is dependent on the common 3D shape $S(\vec{x})$ of the crystal, and each Bragg streak maps a projection through this shape as described by Eqn.~(\ref{eq:projection}). In this situation, the tomographic reconstruction of $S(\vec{x})$ can be achieved from rotation topographs (Sec.~\ref{sec:topo-rotation}) measured from many different RLPs $\vec{G}_\hkl$, as long as each can be normalised by the corresponding structure factor. When the structure factors are not known, a self-consistent reconstruction of all $|F_\hkl|^2$ and $S(\vec{x})$ can be obtained if the crystal volume is fully captured in each topograph. In this case, the integral of a projection of $S(\vec{x})$ is equal to the total diffracting volume of the crystal.

\begin{figure}[tbp]
    \centering
    \includegraphics[width=0.7\linewidth]{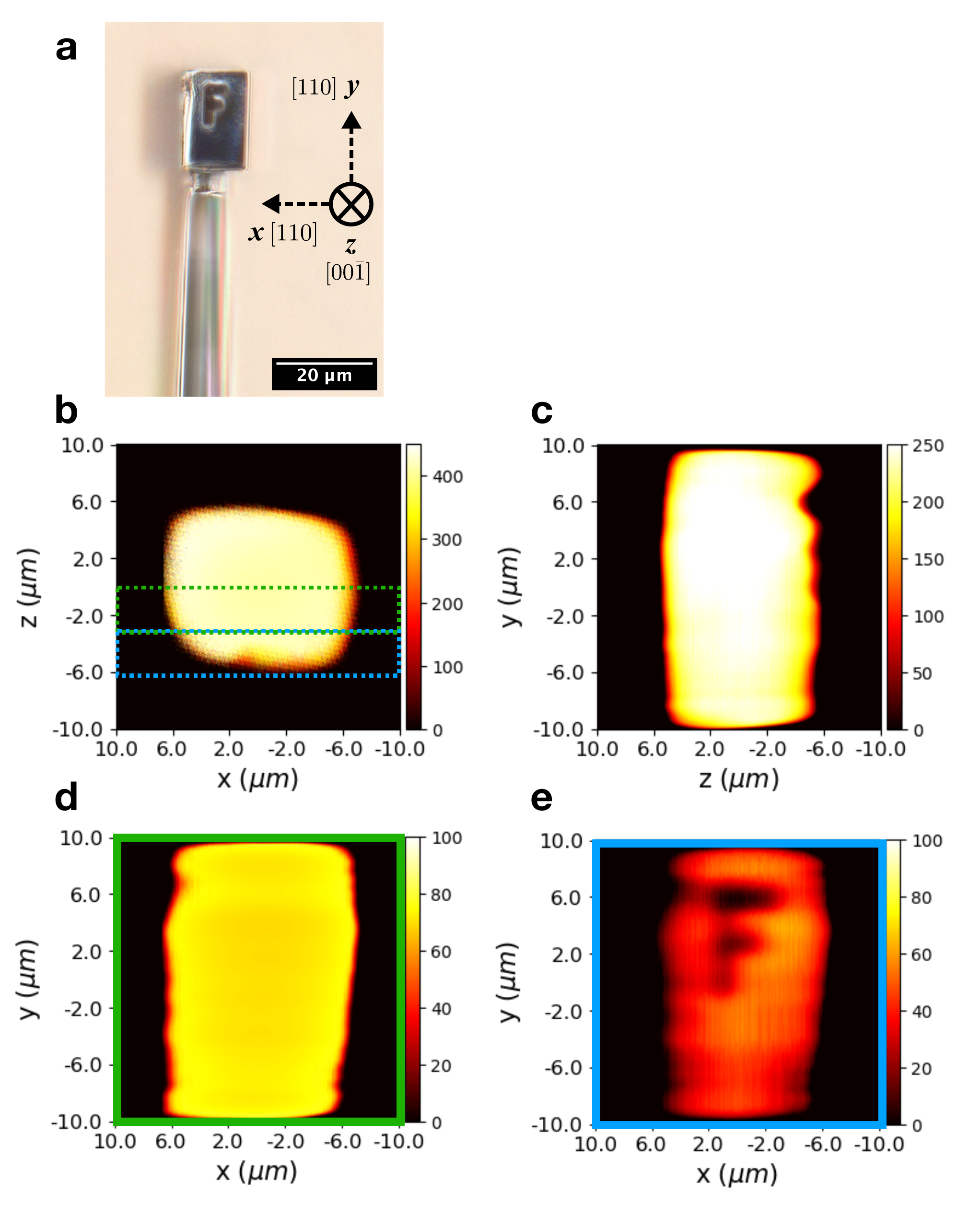}
    \caption{3D tomo-topogram of a single bulk crystal. (a) SEM image of the Si single crystal sample. The crystal is an approximate cubiod with dimensions $10\times20\times \SI{10}{\um\cubed}$ along the $x$, $y$ and $z$ axes, respectively. A letter ``F'' is carved into one surface of the  crystal. (b) \& (c) Views of the crystal shape $S(\vec{x})$ as projected along the $y$ and $x$ axes, respectively. (d) \& (e) Views of slabs of $S(\vec{x})$ indicated by green and blue dashed boxes in (b), projected along the along $z$ axis. The ``F'' letter feature occurs the surface layer and has a depth of about \SI{2}{\um}.}
    \label{fig:si_F_cube_topo}
\end{figure}

In this way, a tomographic reconstruction can be made even when rotating the crystal around a single axis, perpendicular to the optical axis. 
To confirm this, we carried out a measurement on another well-characterised Si crystal shown in Fig.~\ref{fig:si_F_cube_topo} (a). This was prepared by focused ion-beam milling  a crystal to a size of $12 \times 10\times \SI{20}{\micro\meter\cubed}$ with a letter ``F'' milled into one surface to a depth of about \SI{2}{\um}. The crystal was mounted on a pin with the $(1\bar{1}0)$ direction aligned vertically, perpendicular to the rotation axis (parallel to the $y$ axis), and with the $(001)$ direction along the normal vector of the surface with the letter ``F'' ($-z$ axis). The crystal was placed at a defocus of \SI{940}{\um}, giving a magnification of about 174 at a detector distance of \SI{16.4}{\cm}. Diffraction was recorded as the crystal was rotated over a range of \SI{180}{\degree} with a step size of \SI{0.02}{\degree}. The highest order reflection was $531$, at a resolution of \SI{0.92}{\per\angstrom}. In total, 243 undistorted rotation topographs were assembled from the rotation series, and 83 full topographs of high quality were selected, corresponding to 82 views through the crystal with an average angular separation of \SI{2.2}{\degree}. Each topograph was normalized by the factors $L$, $W$, $P$ (the polarisation factor), and by the known structure factors of Si. 

The 3D reconstruction of $S(\vec{x})$ shown in Fig.~\ref{fig:si_F_cube_topo} was achieved using the simultaneous iterative reconstruction technique (SIRT), as detailed in Sec.~\ref{sec:methods-topo}. Calculated re-projections ${\cal{P}}_S$ from the volume in two orthogonal directions are shown in Fig.~\ref{fig:si_F_cube_topo}~(b) and (c). Projections along thin slabs depicted by the green and blue dashed lines in Fig.~\ref{fig:si_F_cube_topo}~(b) are displayed in Figs.~\ref{fig:si_F_cube_topo}~(d) and (e). The slab near the surface of the carved face clearly shows the ``F'' feature, and the spatial resolution can be estimated from this to be about \SI{0.5}{\um}, which is consistent with the Crowther estimate~\cite{Crowther:1970} given by the object diameter multiplied by the average angular separation of views. This test confirms that topographs obtained from various Bragg reflections do provide projection images of the crystal as viewed along the $\kin$ direction. 


\subsection{Structure factors from rotation measurements}
\label{sec:structure-factors}
In a single-crystal X-ray diffraction experiment, structure factor estimates are usually obtained by integrating the counts in Bragg peaks as the crystal is rotated to sweep the Ewald sphere through each RLP volume. The measured intensities may be dependent on the crystal morphology, such as crystal defects, polycrystallinity or twinning, as well as by the absorption of the incident and diffracted beams in the crystal volume. Methods to account for these factors include comparing the intensities of symmetry-related reflections~\cite{Walker:1983} or acquiring additional tomographic measurements of the sample density (but not necessarily of the crystallinity)~\cite{Lu:2024,Polikarpov:2019}. 
When the crystal is placed out of focus and fully illuminated by the X-ray beam, then structure factors can be estimated by integrating the counts in each topograph, giving the potential to avoid including parts of the crystal that may be defective or strained, or even to obtain structure factors as a function of position in a crystal. The complete tomographic information about the crystal, as described in Sec.~\ref{sec:tomo-topo}, can also be used to correct the structure factors for absorption and to accurately estimate the X-ray dose accumulated in a measurement.
From Eqn.~(\ref{eq:2D-topo}),
\begin{equation}
    |F_\hkl|^2 = \frac{1}{c \,P\,L\,W\,V}\iint I_\hkl(\phi_x, \phi_y)/|A(\phi_x, \phi_y)|^2 \,\mathrm{d}\phi_x \,\mathrm{d}\phi_y 
    \label{eq:integration-F}
\end{equation}
where $V$ is the crystal volume that contributes to the measurement. For a rotation series with a fine step size and the crystal fully illuminated by the defocused beam, $V$ is equal to the volume of the entire crystal for all orientations so the integrations of Eqn.~(\ref{eq:integration-F}) can be carried out by summing over all pixels of each topograph. 

To evaluate the validity of Eqn.~\ref{eq:integration-F} (and in particular the factors $L$ and $W$), and to evaluate the accuracies of the resulting structure factor estimates, we carried out the analysis of the same CBXD dataset of the small cuboid Si crystal as used to generate the tomo-topograph shown in Fig.~\ref{fig:si_F_cube_topo}, consisting fo 9000 CBXD patterns with a step size of \SI{0.02}{\degree}. 
Indexing of the rotation dataset of Si could be performed easily by inspection. 

\begin{figure}[!tb]
    \centering
    \includegraphics[width=1\linewidth]{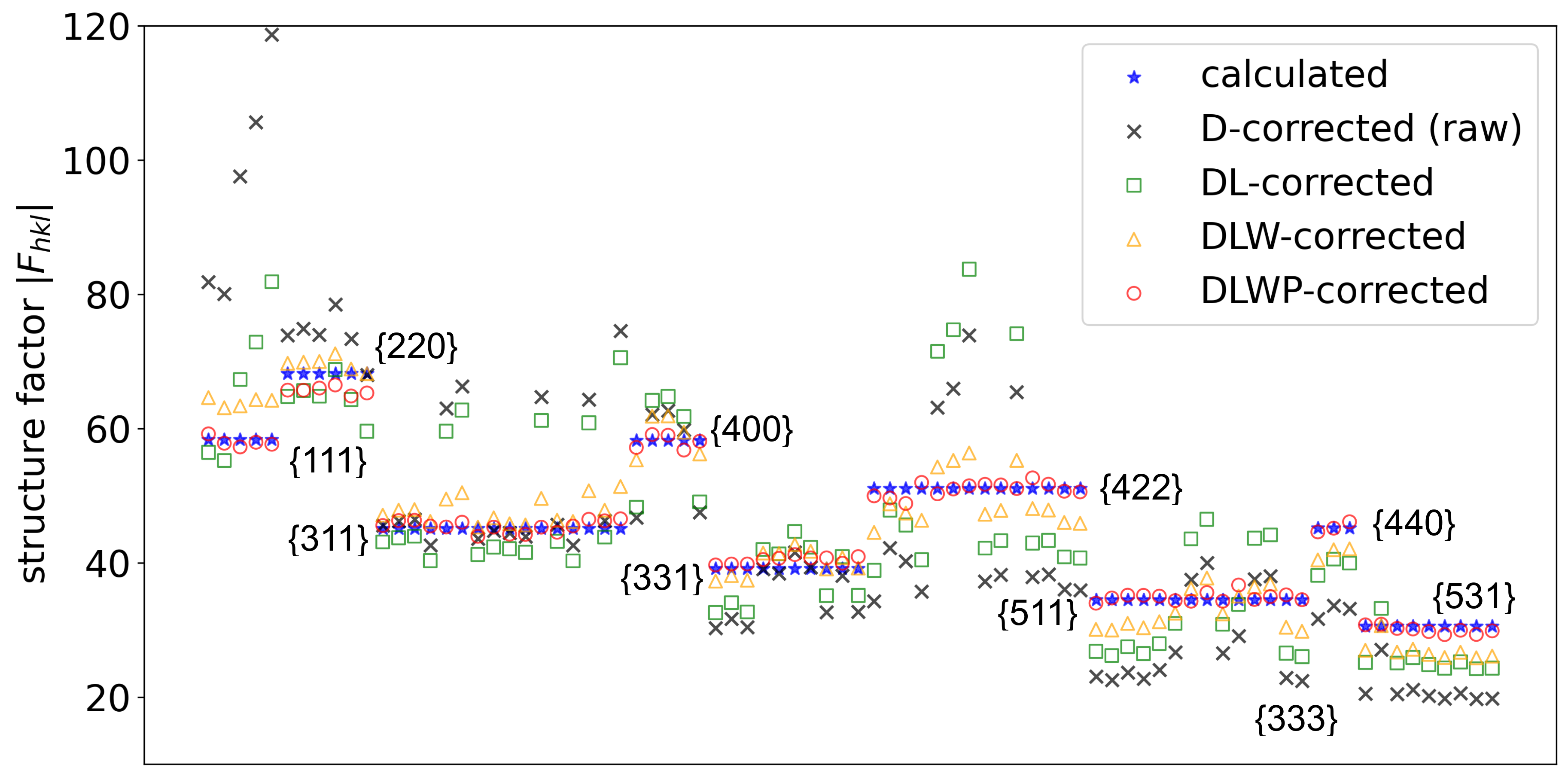}
    \caption{Validation of the intensity factors for CBXD. Structure factor estimates $|F_\hkl|$ of the single crystal silicon ``F'' cuboid sample from a CBXD rotation measurement compared with calculated structure factors (blue stars). The corrections applied are $D$ (black crosses), $D \,L$ (green squares), $D\,L\,W$ (yellow triangles), and $D\,L\,W\,P$ (red circles). The reflections are grouped according to the space group symmetry $Fd\bar{3}m$ of the structure. The structure factors are plotted in units of electrons.}
    \label{fig:corrected_SF}
\end{figure}

Only reflections that gave topographs of the full crystal were used in the  analysis. Each topograph, after normalising by $|A(\phi_x,\phi_y|^2$, provides pixel values from which the mean and standard deviation can be determined of the estimates of the structure factors $|F_\hkl|$. These are plotted as black crosses in Fig.~\ref{fig:corrected_SF}, with only a correction to account for the change in detector quantum efficiency as a function of the scattering angle $2\theta$, given by
\begin{equation}
\label{eqn:App_DQE}
D = 1 - \exp\left(-\frac{\mu_\mathrm{det}\,t_s}{\cos{2\theta}}\right)
\end{equation}
where $t_s$ is the thickness of the sensor material of the detector and $\mu_\mathrm{det}$ is the linear attenuation coefficient of the material. The efficiency increases as the apparent thickness of the sensor material, oriented normal to the optical axis, increases for rays impinging at an angle $2\theta$. Standard crystallography data collection and analysis programs usually account for this factor.

Calculated structure factors of Si were obtained from tabulated values of atomic scattering factors~\cite{Int-tables-SF} after applying a Debye-Waller factor, plotted as blue stars in Fig.~\ref{fig:corrected_SF}. These can be compared in the plot with the values of $\sqrt{\bar{I}_\hkl}$, without correction by the factors $L$, $W$, or $P$ (black crosses). The structure factors are grouped in the plot by symmetry-equivalent reflections. That is, no averaging over symmetry-related reflections was performed. It is seen that there is a wide variation of the uncorrected values within each group of symmetry-equivalent reflections, and the values do not correlate well with the known structure factors, as quantified in Table~\ref{tab:PS_correction}. 

Normalising by the Lorenz factor $L=1/\sin 2\theta_B$ improves the estimated structure factors, shown as green squares in Fig.~\ref{fig:corrected_SF}. However, since this factor is the same for each reflection in a symmetry-equivalent family, it does not improve the variation within each family. This is achieved with the correction of $W= 1/\cos \xi$ to account for the apparent angular velocity of the Bragg streaks.
The yellow triangles in Fig.~\ref{fig:corrected_SF} give the structure factors normalised by both the $L$ and $W$ factors. The final correction by the factor
\begin{equation}
\label{eqn:App_Polarisation}
P = \cos^{2}{2\theta_B}\cos^{2}{\xi} + \sin^{2}{\xi},
\end{equation}
to account for the horizontal polarisation of the incident beam, provides an even better agreement with calculated Si structure factors as shown by the red circles in the figure and as can be seen from Table~\ref{tab:PS_correction}.

Although not a concern for this Si crystal, in larger crystals, X-rays are partially absorbed in the crystal, reducing the measured diffraction counts. The transmission through the crystal for a particular reflection can be calculated by determining all possible paths through the crystal for rays incident at $\kin$ and which scatter into the scattering direction of $\kout$. This can done for a single, perfect crystal, where the density distribution is given by $S(\vec{x})$. Assuming kinematical diffraction, scattering only occurs once and hence along the path of the incident beam, indicated by blue in Fig.~\ref{fig:CBXD_geometry}~(f). The transmission for a given pair of wave-vectors $\kin$ and $\kout$ can be calculated for all paths $L_\mathrm{in}(s_0) + L_\mathrm{out}(s_0)$ intersecting the crystal shape $S(\vec{x})$ where the scattering point $s_0$ ranges from 0 to the crystal thickness $t$, as
\begin{equation}
\label{eq:A-absorption}
T(\kin;\kout) = \frac{1}{t} \int_{0}^{t}\exp\left[-\mu_\mathrm{xtal} \,\{{L_\mathrm{in}(s_0)+L_\mathrm{out}(s_0)\}}  \right]\,\mathrm{d}s_0,
\end{equation}
where $\mu_\mathrm{xtal}$ is the linear absorption coefficient of the crystal material. The transmission can be calculated for all $\kin(\chi)$ vectors of the Bragg streak or for all $\kin(\phi_x,\phi_y)$ vectors of a topograph. This calculation can also be used to accurately compute the dose to the crystal.
The longest path length through the Si crystal was less than \SI{25}{\um}, and we find that absorption factor is generally not less than 0.99. 

When all corrections are incorporated, we find that the measurements gave an error of about \SI{2}{\percent}, shown as red circles in Fig.~\ref{fig:corrected_SF}. The residual error might be explained by dynamical diffraction effects and variability of the incident flux. The analysis does show that the incorporation of every correction factor brings the estimated structure factor closer to the known value, validating the derivation and need for each.

\begin{table}[!tb]
    \centering
    \begin{tabular}{c|cccc}
        \hline
         Correction &  $|A|^2D$ (Raw) &  $|A|^2DL$&  $|A|^2DLW$ & $|A|^2DLWP$\\
         \hline
         Pearson correlation coefficient&  0.7813&  0.7738&  0.9648& 0.9951\\
         $R$ factor&  0.2316&  0.1676&  0.0685& 0.0200\\
         \hline
    \end{tabular}
    \caption{Pearson correlation coefficients and $R$ factor between experimental values and calculated values of structure factor moduli of silicon crystal to a resolution of 0.92 \AA, with various corrections applied.}
    \label{tab:PS_correction}
\end{table}

\section{Example: Vitamin B$_{12}$ convergent-beam diffraction}
\label{sec:B12}
\begin{figure}[!tb]
    \centering
    \includegraphics[width=\linewidth]{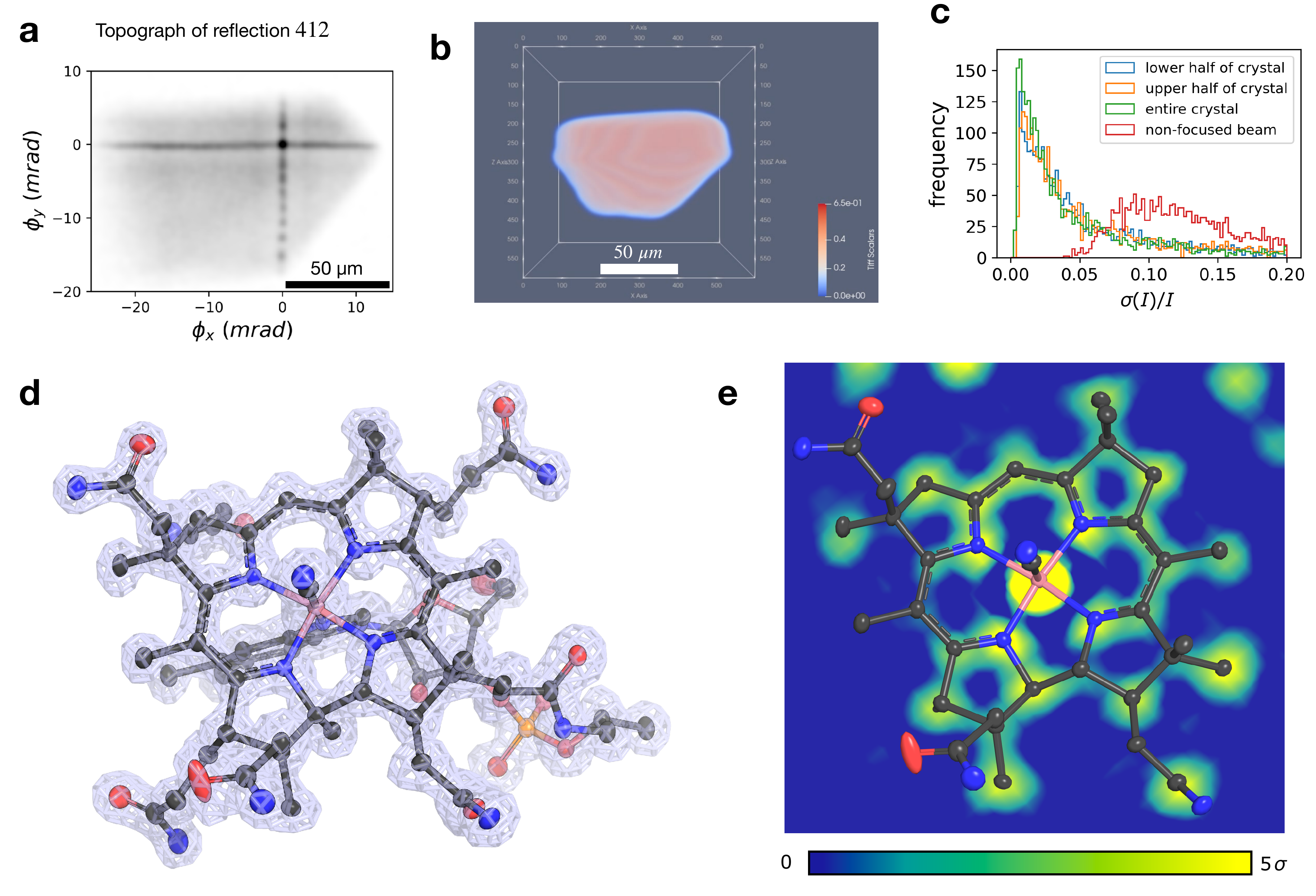}
    \caption{Crystal and molecular structure of the vitamin B$_{12}$ crystal. (a) Rotation topograph of the 412 reflection of the vitamin B$_{12}$ crystal used in Fig.~\ref{fig:B12-snapshot} (b). Rendering of the crystal volume $S(\vec{x})$ obtained from all topographs. (c) Histograms of relative errors $\sigma(\bar{I})/\bar{I}$ of reflections obtained from the full crystal, the top and bottom halves, and from the collimated beam. (d) Structure and $\sigma$-weighted $2F_\mathrm{o}-F_\mathrm{c}$ electron density map (mesh) of vitamin B$_{12}$ obtained from the entire crystal volume. (e) Map of the electron density in a slice through the corrin ring of vitamin B$_{12}$.}
    \label{fig:B12-structure}
\end{figure}

Following the approaches of Sec.~\ref{sec:rotation-CBXD}, topographs and structure factors of a vitamin B$_{12}$ crystal were obtained from a rotation series. The vitamin B$_{12}$ crystal was about \SI{140}{\um} wide and was placed out of focus by \SI{3.5}{\mm} to ensure that it was fully illuminated throughout the entire rotation scan (see Methods, Sec.~\ref{sec:methods-instrument}). The crystal was rotated about the vertical ($y$) axis with a step size of \SI{0.05}{\degree} and an exposure time per step of \SI{0.5}{\second}. The static diffraction pattern shown in Fig.~\ref{fig:B12-snapshot} (d) is one exposure of the rotation series. The result of a rotation topograph of the 412 reflection, formed over a rotation range of \SI{2.5}{\degree} and corrected for the aperture function, is shown in Fig.~\ref{fig:B12-structure}~(a), again in good agreement with the optical image of Fig.~\ref{fig:B12-snapshot} (f). The cross pattern of higher counts along the lines $\phi_x = 0$ and $\phi_y = 0$, due to diffraction from the two orthogonal line focii, aid indexing, as described in Sec.~\ref{sec:methods-instrument}. These features were excluded from the integration of counts in the topograph since the illumination volume of the crystal for these beams is not well known. 

A model of the crystal shape $S(\vec{x})$ was constructed from undistorted full topographs obtained at 36 crystal orientations (Sec.~\ref{sec:tomo-topo}), as shown in Fig.~\ref{fig:B12-structure}~(b). Structure factors were then obtained by integrating  those topographs that displayed the entire crystal. 
However, for a rotation series about a single axis perpendicular to the optical axis, there are many reflections close to that axis that only produce partial topographs, and which otherwise would require rotation about a second axis to complete. This may also be the case if the scan is performed in blocks of rotations to reduce the total number of recorded diffraction patterns (to save time or reduce dose). This real-space partiality problem was addressed by fitting a scale factor to match the partial topograph to a re-projection of the tomographic reconstruction of $S(\vec{x})$. 
In this way, a complete set of structure factors was obtained to a resolution of $\lambda/(2\sin\theta) = \SI{1.2}{\angstrom}$.

\begin{figure}[!tb]
    \centering
    \includegraphics[width=0.6\linewidth]{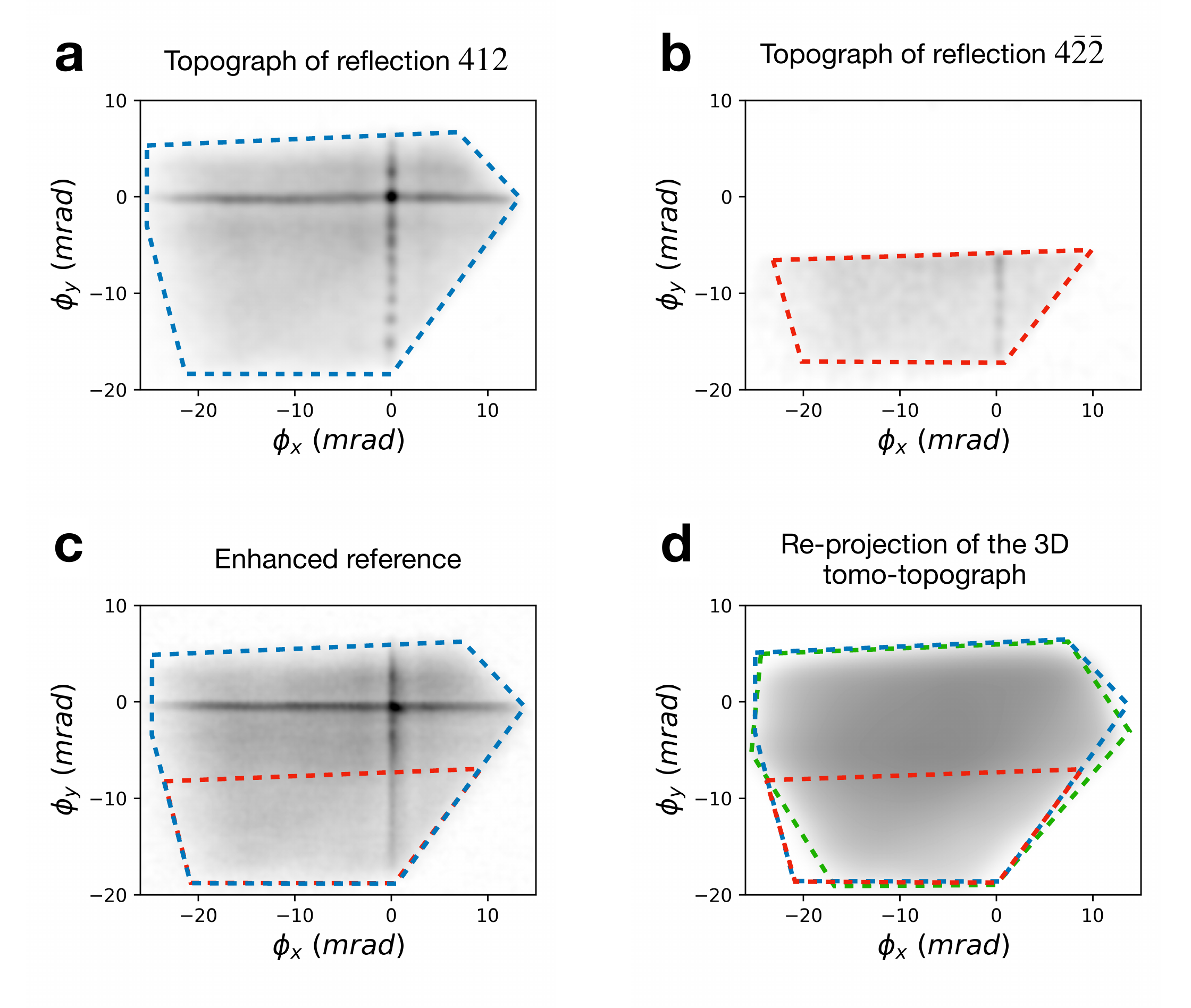}
    \caption{Estimation of structure factors from partial topographs. (a) A full topograph for the vitamin $B_{12}$ crystal of Fig.~\ref{fig:B12-snapshot}, formed from reflection $412$. (b) A partial topograph from reflection $4\bar{2}\bar{2}$. (c) The enhanced reference obtained by summing multiple full topographs of similar orientations. (d) Re-projection ${\cal{P}}_S(\phi_x,\phi_y)$ at the crystal orientation $\vec{\mathbf{\omega}}$ giving the $4\bar{2}\bar{2}$ reflection. 
    The crystal boundaries (indicated by the colored dashed polygons) were determined for each full or partial topograph, and the topograph was masked in regions outside the boundary. Only the Bragg signal within the red boundary was used to estimate the structure factor by scaling each topograph to either the reference or the re-projection.}
    \label{fig:topo_scaling}
\end{figure}

An example of the fitting of a partial topograph is shown in Fig.~\ref{fig:topo_scaling}. The measured topograph of the $4\bar{2}\bar{2}$ reflection is shown in Fig.~\ref{fig:topo_scaling}~(a), and the calculated reprojected topograph ${{\cal{P}}}_S$ in the $\kin$ direction for that $4\bar{2}\bar{2}$ reflection is given in (d). This reflection was relatively weak, and the reprojected topograph enables a mask to be defined for the region for integrating counts, as shown in Fig.~\ref{fig:topo_scaling}~(c). The sum of counts over this masked region is then multiplied by the ratio of the integrated partial reprojected view (d) to the full reprojected view in (b). A similar fitting procedure can also be used on complete topographs to improve the estimate of $|F_\hkl|^2$, and on sparse topographs that were recorded with a rotational step size much larger than the Bragg streak width $\Delta \theta$ (such as illustrated in Fig.~\ref{fig:si_wedge_hole_topo}~(f)).

The tomograph of the crystal shape function was also used to correct for absorption in the crystal as described in Sec.~\ref{sec:structure-factors}. This calculation gave a correction of 0.95 for this crystal of maximum diameter of \SI{140}{\um}. For this sized crystal there was no need to consider dynamical diffraction effects. The same ray tracing to compute the absorption also enables an accurate estimation of the dose in the crystal, giving an estimate of \SI{42}{\kilo\gray} for an incident fluence of \SI{5e9}{\photon\per\second} in the focused beam.

Since structure factors are determined from topographs of the crystal, in this case the relative standard error of the integrated intensities, $\sigma(\bar{I})/\bar{I}$, can be derived from the photon counts measured in many different detector pixels. This is done by computing the standard deviation of the values of the topograph, normalised by $S(\vec{x})$ and by the lens aperture function, and excluding regions due to the line foci and direct beam. Histograms of the relative errors, $\sigma(\bar{I})/\bar{I}$, are shown for the vitamin B$_{12}$ crystal in Fig.~\ref{fig:B12-structure} (c). The approach of obtaining integrated intensities and errors by CBXD is compared with that of a conventional rotation dataset with a collimated beam, by plotting the distribution of errors as computed by utilising intensities originating only from the unfocused beam component at $(\phi_x, \phi_y) = (0,0)$. The total counts in the unfocused and focused beams were comparable, but the mean and median errors are larger for the collimated beam.

The structure of vitamin B$_{12}$ obtained from the measured structure factors is displayed in Figs.~\ref{fig:B12-structure} (d) and (e), and data and refinement statistics are given in Table~\ref{tab:B12} (see Sec.~\ref{sec:B12-method}).  The quality of the CBXD reconstruction is restricted by the limited resolution of the diffraction data and possibly also due to background scattering and radiation damage. Nevertheless, the result validates the data collection methodology and analysis. 

\begin{table}[]
    \centering
    \begin{tabular}{llll}
 \hline
   Parameter                                     & Full crystal              & Upper half           & Lower half         \\
 \hline
Empirical formula                                & \ce{C63H87CoN14O33P}  & \ce{C63H87CoN14O33P}  & \ce{C63H87CoN14O33P}  \\
Formula weight (Da)                              & 1658.36          & 1658.36          & 1658.36          \\
Temperature (K)                                  & 293              & 293              & 293              \\
Crystal system                  & Orthorhombic   & Orthorhombic    & Orthorhombic   \\
Space group & \it{P}$2_12_12_1$  & \it{P}$2_12_12_1$  & \it{P}$2_12_12_1$ \\ 
a (Å)                                            & 15.704(16)       & 15.704(16)       & 15.704(16)       \\
b (Å)                                            & 22.15(2)         & 22.15(2)         & 22.15(2)         \\
c (Å)                                            & 24.96(3)         & 24.96(3)         & 24.96(3)         \\
$\alpha$, $\beta$, $\gamma$ ($^\circ$)           & 90, 90, 90       & 90, 90, 90               & 90, 90, 90               \\
Volume (Å$^3$)                                   & 8683(15)         & 8683(15)         & 8683(15)         \\
$Z$                                              & 4                & 4                & 4                \\
Density (g/cm$^3$)                                 & 1.269            & 1.269            & 1.269            \\
Wavelength (Å)                                   & 0.708            & 0.708            & 0.708            \\
Abs. coefficient (mm$^{-1}$)                     & 0.302            & 0.302            & 0.302            \\
Crystal size ($\mu$m$^3$) & 140 $\times$ 90 $\times$ 70 & 140 $\times$ 40 $\times$ 70 & 120 $\times$ 50 $\times$ 70\\

$F$(000)                                           & 3476             & 3476             & 3476             \\
$\theta$ range ($^\circ$)                                      & 1.526--17.355   & 1.526--17.294   & 1.526--17.355   \\
Reflections collected                            & 15605            & 14895            & 16343            \\
Independent reflections                          & 5281             & 5251             & 5305             \\
Observed reflections {[}$I$ \textgreater $2\sigma(I)${]} & 3907             & 3521             & 3581             \\
$R_\mathrm{int}$                                           & 0.1396           & 0.1407           & 0.1502           \\
Goodness-of-fit on $F^2$          & 0.825            & 0.772            & 0.781            \\
R {[}$I > 2\sigma(I)${]}                    & 0.0841           & 0.0823           & 0.0826           \\
$wR$($F^2$) (all data)              & 0.2303           & 0.2295           & 0.2283           \\
Data / restraints / parameters                   & 5281 / 860 / 958 & 5251 / 860 / 958 & 5305 / 860 / 958
    \end{tabular}
    \caption{Data collection and refinement statistics of the vitamin B$_{12}$ crystal.}
    \label{tab:B12}
\end{table}

\subsection{Volume dependent structure factors}
A potential of CBXD is to extract structure factors from different regions of the crystal volume, $S(\vec{x})$, e.g.\ when the occupancy of a binding ligand varies as it diffuses into a protein crystal or to determine the relationship between conformation and crystal strain in a photoreactive crystal. Obtaining the diffraction signal from a sub-volume of the crystal can be described as inverse tomography, equivalent to the task in radiotherapy of concentrating dose in a particular region by tailoring the radiation illumination profile at various orientations of the patient~\cite{Oelfke:1999}. Given $S(\vec{x})$, the contribution of a particular voxel can be projected onto each measurement, but solving for the value of the signal in the voxel requires measurements along many lines of sight. Extracting structure factors from particular regions of a crystal is complicated by the fact that the projection topograph for a particular RLP can only be obtained when it is in the diffracting condition. This necessitates collecting data for crystal rotations around several axes, or by comparing topographs of reflections that are equivalent due to crystal symmetry. Here, we demonstrate the general concept on the vitamin B$_{12}$ dataset by extracting two sets of structure factors from the top and bottom halves of the crystal as it is rotated about the vertical axis. In this rotation series, these regions never occlude each other, enabling every topograph to be separated into its top and bottom contributions. The structure factors and relative errors (Fig.~\ref{fig:B12-structure} (c)) were nearly identical for the two halves, as would be expected for a flawless crystal. Refinement statistics for the two halves are given in Table~\ref{tab:B12}.   



\section{Methods}
\label{sec:methods}
\subsection{Experiment and instrumentation}
\label{sec:methods-instrument}
Experiments were performed at a photon energy of \SI{17.5}{\kilo\eV} (\SI{0.7}{\angstrom} wavelength) at the P11 beamline of the PETRA III synchrotron radiation facility, using an X-ray microscope setup~\cite{Zhang:2024} to mount and align the MLLs and to position and rotate the crystal sample. Diffraction data was recorded on an EIGER X 16M detector with a Si sensor (Dectris) consisting of $4148 \times 4362$ pixels. This detector was mounted on rails and used at a long distance of about \SI{2}{\meter} for aligning and characterising lenses, and then brought to a distance of \SIrange{16}{20}{\cm} for high-angle diffraction measurements. The beam illuminating the lenses was passed through a Si 111 channel-cut monochromator and then broadened to a height and width of more than \SI{5}{\mm} by adjusting the bending of the beamline Kirkpatrick-Baez mirrors. This was done to increase the transverse spatial coherence of the incident beam to match the width of the lens apertures. With full coherence, the lenses will focus to a diffraction-limited spot, but broadening the incident beam reduced the measured counts in the focused beam to about \SI{e10}{\photon\per\second}. In retrospect, it was not necessary to increase the transverse coherence length since the topographs were only recorded at micrometer resolution, and thus it may have been possible to record data about 100 times faster with an illuminating beam of $\SI{0.5}{\mm} \times \SI{0.5}{\mm}$.  

Two pairs of lenses were used for the experiments. All lenses were prepared by masked deposition using magnetron sputtering~\cite{Prasciolu:2015}. Each lens of the pair was mounted on a separate hexapod in the X-ray microscope and then brought into its diffracting condition to give the highest counts into the focused beam as measured on the EIGER detector. The lenses were then overlapped in the beam to produce a two-dimensional focus. From the two astigmatism terms of the wavefront measured by speckle tracking~\cite{ivanov:2022,Dresselhaus:2024}, adjustments were made to ensure the lenses focused in orthogonal directions to a common plane~\cite{Zakharova:2025}. 
For the measurements of the Si crystals, a pair of off-axis MLLs were used with 0.015 NA and focal lengths of \SI{1.15}{\mm} and \SI{1.25}{\mm} at \SI{17.5}{\kilo\eV}. Both lenses of this pair consisted of \num{10855} bi-layers, with layer periods ranging from \SI{8.8}{\nm} to \SI{2.0}{\nm}. With this, each lens aperture was about \SI{35}{\um} high, ranging from \SI{10}{\um} from the optical axis to \SI{45}{\um} from the optical axis. A second pair of MLLs  were used for the measurements of the vitamin B$_{12}$ crystal. These were on-axis lenses with 0.028 NA and focal lengths of \SI{1.25}{\mm} and \SI{1.26}{\mm} at \SI{17.5}{\kilo\eV}, consisting of \num{16965} bi-layers with a minimum period of \SI{2.06}{\nm}. The lens apertures were about \SI{70}{\um} high, centered on the optical axis. 

Multilayer Laue lenses themselves diffract X-rays according to the principles of dynamical diffraction and each lens gives rise to a focused (diffracted) and unfocused (zero order) beam~\cite{Yan:2014}. The pair of lenses thus create a 2D focus, two orthogonal line foci, and a zero order beam. These components are composed of incident wave-vectors respectively given by $\kin(\phi_x, \phi_y)$, $\kin(\phi_x, 0)$, $\kin(0, \phi_y)$, and $\kin(0,0)$ where $\phi_x$ and $\phi_y$ range over $-\alpha$ to $\alpha$. The zero order and line foci can be fully or partially blocked by a combination of an order-sorting aperture upstream of the focus and a complementary central stop and aperture upstream of the lenses. For an off-axis lens, the 2D focused beam diverges from the zero orders and so a crystal that is fully illuminated by that beam does not intersect the line foci or the collimated beam. For a pair of on-axis lenses, however, these will intersect with the crystal and give rise to additional components of diffraction. For example, for the vertical line focus, additional diffraction intensity will be observed in the direction $\kin(\phi_x,0)+\vec{G}$ if the deficiency line crosses the $\phi_x$ axis. These components, including diffraction by the zero order $\kin(0,0)$ beam, are very apparent in rotation topographs, as seen in Fig.~\ref{fig:B12-structure} (a). They directly provide the $(\phi_x,\phi_y)$ coordinates.

The Si crystals were mounted on the tops of glass pins, while the vitamin B$_{12}$ crystal was mounted in a loop, contained within a sealed kapton tube. These were placed on a sample stage with $x$-$y$-$z$ positioning and rotation around the vertical ($y$) axis. A square order-sorting aperture of \SI{20}{\um} width was placed between the lenses and sample, close to the focal plane, to cut most of the unfocused orders of the lenses, and to block diffuse scattering on the detector from the lens materials. The crystals were placed downstream of focus by a sufficient distance to fully illuminate them in the diverging beam. All measurements were made at room temperature.

\subsection{Analysis of diffraction data}
\label{sec:methods-diffraction}
The CBXD data in a rotation scan were processed via the steps of background subtraction of diffraction patterns and masking of improper pixel values, streak detection, indexing, intensity integration, and merging of intensities.

\subsubsection{Pre-processing of diffraction frames} 
\label{sec:methods-pre-processing}
Raw diffraction frames were first corrected for the detector quantum efficiency, $D$ (Eqn.~\ref{eqn:App_DQE}). In addition to the desired CBXD signal of Bragg streaks, diffraction patterns contain a diffuse background due to scattering and fluorescence from any matter along the beam path (including air, solvent, sample support), which was approximately constant throughout the scan. This background is best computed after excluding the Bragg streaks from the analysis, but to find all Bragg streaks we must first subtract the background. Our strategy was to subtract an initial estimate of the background at each detector pixel to improve the detection of Bragg streaks, and then re-estimate the background only after the positions of predicted Bragg streaks matched observations. This initial estimate of the background was obtained from the median value of counts in the pixel as a function of frame number. In this process, some pixels were identified as bad if they gave consistently low values (“dead” pixels that are unresponsive to photons) or consistently high values (“hot” pixels). These pixels were masked from the analysis. Prior to subtracting the background estimate, pixels with saturated counts were identified.

\subsubsection{Streak detection}
Streaks were identified as connected regions of pixels that exceed a particular intensity threshold. They were detected by creating a binary array equal to 1 in pixels that exceed the threshold.  A group of pixels was considered as connected if they join at shared edges or corners. All such self-connected regions were characterised by their pixel count, coordinates, length, width, intensity-weighted centroid, orientation, and aspect ratio. Bragg streaks were then selected as regions exceeding a given number of connected pixels. This bound, and the intensity threshold was optimised by inspection of the patterns.

\subsubsection{Indexing---initial estimation}
Indexing requires determining both the RLP associated with each Bragg streak and the incident $\kin$ deficiency line. One approach to index a rotation series is to initially assume that the centroid of each distorted topograph, located at $\vec{k}_{\mathrm{out},c}$, is generated by the central ray of the lens aperture (the chief ray),  $\kin =\kin(0,0)$, so that $\vec{G}=\vec{k}_{\mathrm{out},c} - \kin(0,0)$. This would be the case if both the crystal and the rotation axis of the crystal were centered on the beam diverging from the focus, an assumption that is not necessarily true, as can be seen from Fig.~\ref{fig:B12-snapshot}~(e). Nevertheless, the error in locating $\vec{G}$ this way can not be more than the beam convergence angle $\alpha$, which is usually smaller than the angle between neighboring RLPs. Using the \textit{SPIND} software~\cite{Li:2019b}, the lattice points $\vec{G}$ found in this way are then placed as voxels in a 3D reciprocal space array which is Fourier transformed to obtain a first estimate of the lattice constants of the crystal and its orientation matrix. We observed that the autocorrelation of the reciprocal lattice array obtained by the Fourier transform extended over several real-space lattice positions—enough to give a good estimate of the lattice parameters (if not known \emph{a priori}) and the orientation matrix.

This approach to obtain an initial estimate of the the lattice parameters and orientation matrix can also be used for data collected over small rotation ranges, or even a single pattern. The centroid of the Bragg streak or partial topograph is then be used to estimate $\vec{G}$. This approach is aided by the fact that many more reflections are excited in a CBXD pattern than for a collimated beam. 

As mentioned in Sec.~\ref{sec:methods-instrument}, on-axis MLLs provide the $(\phi_x, \phi_y)$ coordinates for each topograph. 
In that case, the location of the $\kout$ vector corresponding to the $\kin(0,0)$ illumination of each topograph, and the crystal rotation $\vec{\omega}$ for that diffracting condition, can then be used in any indexing algorithm for conventional rotational crystallography, such as \textit{MOSFLM}~\cite{Leslie:2006} or \textit{XDS}~\cite{Kabsch:2010} for macromolecular crystals. For snapshot diffraction patterns, the orthogonal fiducial diffraction also constrains the $\kin$ coordinates of the streaks. One high-intensity pixel in a Bragg streak indicates it crosses either one of the $\phi_x$ or $\phi_y$ axes, and two high-intensity pixels provide two discrete options.

\subsubsection{Indexing---refinement}
After obtaining an initial estimate of the lattice parameters and orientation matrix these quantities were then further refined by comparing the positions of the observed positions $\kout^\mathrm{obs}$ in every pattern with Bragg streaks $\kout^\mathrm{calc}$ calculated from Eqn.~(\ref{eq:circle}) for lattice parameters $p_\mathrm{x}$, the orientation matrix $\mathbf{R}$, and parameters that define the experiment geometry, $p_\mathrm{g}$. The refinement of each pattern minimises a loss function with respect to the parameters, defined as
\begin{equation}
         {\cal{L}} = \frac{1}{N_s}\sum_{i}^{N_s}\left| \vec{k}_{\mathrm{out},i}^\mathrm{obs}- \vec{k}_{\mathrm{out},i}^\mathrm{calc}\left(p_\mathrm{g},p_\mathrm{x},\mathbf{R} \right)\right|
\end{equation}
where the index $i$ denotes each Bragg streak and $N_s$ is the total number of observed streaks in the pattern. The geometrical parameters include the distances of the detector and crystal from the beam focus, the coordinate and tilt of the chief ray of the lens, and the wavelength. The minimisation of $\cal{L}$ was first carried out using the differential evolution (DE) algorithm~\cite{Storn:1997,Das:2016} and followed with the application of a gradient descent method. When the on-axis MLL was used, the constraints of the $\kin$ coordinates, such as the NA and $\kin(0,0)$,  were applied in a second round of refinement. This usually brought the predicted positions in agreement with the observations to pixel precision. 

\subsubsection{Generation of topographs}
\label{sec:methods-topo}
After the refinement of the geometry and crystal parameters, the background in each diffraction pattern was recalculated, using both detected and predicted Bragg streak locations to define a mask for each pattern that excludes all Bragg streaks. Background counts for each pixel in a pattern were then estimated from the average of unmasked pixel values at that same location for the 50 frames closest in rotation angle to that pattern in the rotation series. This provides a much more accurate estimate of the background, which was subtracted from the raw patterns. The background-corrected pixel values within the predicted Bragg streak masks were then mapped to the $\kin(\phi_x,\phi_y)$ coordinates, and normalised by the measured aperture function $|A(\kin)|^2$ to form real-space undistorted topographs (Sec.~\ref{sec:topo-rotation}). Prior to this normalisation, the aperture function was smoothed to account for inaccuracies in the mapping to $\kin$ coordinates.

We make the approximation that each topograph is the projection through the crystal shape in a fixed orientation as described by Eqn.~\ref{eq:2D-topo}. The 3D volume of the crystal shape was computed tomographically from a number of views $N$ that was usually chosen to be less than the number of observed Bragg streaks, typically $N = 36$. For each orientation bin, an enhanced topograph was obtained from the sum of all topographs obtained from Bragg streaks observed within that orientation range and then normalised by the total integrated counts. The normalised topographs were then combined tomographically to form the 3D image $S(\vec{x})$ of the crystal. This was done using the simultaneous iterative reconstruction technique (SIRT)~\cite{Kak:1988} implemented in the \textit{TomoPy} python package~\cite{Gursoy:2014}. 

\subsubsection{Calculation of structure factors}
\label{sec:methods-structure-factors}
The tomo-topograph serves as a complete model of the crystal, $S(\vec{x})$, and is used to account for the dependence of the diffraction counts in Bragg streaks and rotation topographs on the projected thickness of the crystal. For each RLP, a re-projected 2D map $\mathcal{P}_S(\hkl)$ was computed from $S(\vec{x})$. Next, estimates integrated intensities $\bar{I}_\hkl$ were obtained from full or partial topographs $I_\hkl(\phi_x,\phi_y)$ by performing a pixel-wise division of $I_\hkl(\phi_x,\phi_y)$ by $\mathcal{P}_S(\hkl)$. Where warranted, the topographs can also be divided, pixel-wise, by the transmission factor of Eqn.~\ref{eq:A-absorption} to correct for absorption in the crystal. Each pixel value of this map should be proportional to the squared structure factor, and so the entire map provides many independent measures of that quantity. The mean of this map, $\bar{I}_\hkl$, accounts for the measurement of an incomplete topograph. The median can also be computed as a more robust estimate that avoids the influence of outliers. 

The integrated intensities $\bar{I}_\hkl$ were finally corrected by the factors $P$ (Eqn.~(\ref{eqn:App_Polarisation})), $L$ (Eqn.~(\ref{eq:rocking})), and $W = 1/\cos \xi$. For higher precision, these corrections can instead be first performed pixel-wise on the topographs $I_\hkl(\phi_x,\phi_y)$. 
The errors in the corrected integrated intensities were evaluated as the standard deviation of the pixel-wise division of $I_\hkl(\phi_x,\phi_y)$ by $\mathcal{P}_S(\hkl)$. Errors were estimated for individual reflection measurements, and also by comparing with the values of symmetry-related reflections in the merging stage. 


\subsection{Structure determination of vitamin B$_{12}$}
\label{sec:B12-method}
The structure was solved using \textit{ShelXt}~\cite{Sheldrick:2015} using the known space group and unit-cell parameters, as well as the atom types that form the molecule. The initial structure was inspected using \textit{ShelXle}~\cite{Hubschle:2011} and \textit{Olex2}~\cite{Dolomanov:2009} and corrected to match the known composition of cyanocobalamine. The H atoms were located according to their heavier bonding partner atoms (``riding-H'') with constraints on all bond lengths and angles to ensure they are chemically reasonable.  The model was further refined using \textit{ShelXl}~\cite{Sheldrick:2015L} where non-H atoms were refined anisotropically. The final refinement parameters can be assessed by the respective .res, .ins and .cif files attached as supplemental files. The images of Figs.~\ref{fig:B12-structure} (d) and (e) were created using \textit{PyMOL}~\cite{pymol}. The $\sigma$-weighted $2F_\mathrm{o}- F_\mathrm{c}$ map of vitamin B$_{12}$ is shown contoured at $1.5 \,\sigma$ and carved at \SI{1.2}{\angstrom} around the atoms. The \ce{H2O} molecules in the structure, as well as the Hydrogen atoms of the vitamin B$_{12}$, are hidden. For these maps , ccp4-maps were created from .fcf files in \textit{Coot}~\cite{Emsley:2010} and, likewise, .res-files were exported to .pdb-files for better handling in \textit{PyMOL}. 


\section{Discussion and conclusions}\label{sec:discussion}
Using newly available multilayer Laue lenses, CBXD combines X-ray microscopy with crystallography. The convergence angles of MLLs can be as high as several degrees, creating a focused probe that is only several nanometers in size, comparable to the unit cell size of a protein crystal~\cite{Dresselhaus:2024}. Even when illuminated by this small focus, the diffraction of a 3D crystal consists of Bragg reflections, essentially due to the periodicity apparent along the beam axis. These reflections occur for RLPs that are located in the volume of reciprocal space swept by Ewald spheres generated from all incident wave-vectors supplied by the lens. The Bragg reflections take the form of streaks, each formed by diffraction of the corresponding deficiency line of incident wave-vectors from the lens that together lie on the Kossel cone~\cite{Frank:1972}. The length of the Bragg streaks is thus determined by the lens convergence angle and the short direction of the streak provides the full rocking curve of the reflection. 

For a crystal in focus, all Bragg streaks originate from a common focal volume that intersects the crystal. Images could therefore be constructed from various diffraction signals by rastering the crystal through the beam, as in the approach of Bragg ptychography~\cite{Hruszkewycz:2017}. In practice, using synchrotron radiation, the tightly focused beam quickly damages molecular crystals, and here we explored a different approach of generating magnified topographs (maps of the diffraction efficiency) with the crystal placed out of focus. The resolution of these topographs is set by the pixel size of the detector and the magnification as set by the relative distance of the detector or the crystal to the focus. Since the detector is usually set to a short distance (tens of centimeters) as needed to measure Bragg reflections to a high scattering angle, the magnification may depend on the crystal size (placed as close to the focus where it is still fully illuminated by the diverging beam). The topographs of the crystal will then be sampled according to the number of pixels that the diverging beam covers at the detector. For a semi-convergence angle $\alpha = \SI{1.6}{\degree}$ and detector distance of \SI{20}{\cm}, as used here, there were 74 pixels (\SI{75}{\micro\meter} wide) across the beam, and so sub-micron sampling is achieved for crystals smaller than \SI{74}{\micro\meter}. Larger crystals could be analysed at this resolution by additionally translating the defocused crystal to form a number of rotation topographs that are then stitched together. 

A two-dimensional topograph of the defocused crystal can be formed from a single static diffraction pattern that has been accurately indexed to map streaks back to their incident deficiency lines. This mapping is usually dense enough to provide a detailed image of the diffraction efficiency after normalizing each streak by the square modulus of the structure factor, as shown in Fig.~\ref{fig:bootstrap}, which then allows a better estimation of the structure factors---essentially solving a real-space partiality problem. This analysis is valuable for serial  crystallography~\cite{Chapman:2025}, but in rotational crystallography, full topographs can be obtained from each Bragg streak as its deficiency line sweeps across the illuminated crystal. In this case, lattice strain can be discerned from the distortion of the Bragg streaks and, using information from many reflections, it may be possible to generate maps of the strain tensor. Here, we demonstrated reconstructing the 3D image of the diffraction efficiency by a tomographic analysis of topographs from a Si crystal and a vitamin B$_{12}$ crystal.

In a single-crystal X-ray diffraction experiment, structure factor estimates are usually obtained by integrating the counts in Bragg peaks as the crystal is rotated to sweep the Ewald sphere through each RLP volume. The measured intensities may be dependent on the crystal morphology, such as crystal defects, twinning, or polycrystallinity~\cite{Vlahakis:2024}, and by the absorption of the incident and diffracted beams in the crystal volume. Methods to account for these factors include comparing the intensities of symmetry-related reflections~\cite{Walker:1983} or acquiring additional tomographic measurements of the sample density (but not necessarily of the crystallinity)~\cite{Lu:2024,Polikarpov:2019}. The topographic analysis obtained by CBXD can be used to solve for a self consistent set of structure factors while accounting for crystal shape and order. The analysis pipeline was validated by obtaining structure factors of Si to \SI{2}{\percent} of known values and by solving the structure of vitamin B$_{12}$ using direct methods. 

The approach can be used to extract structure factors from particular regions of the crystal. Here, we obtained the structure from two halves of the vitamin B$_{12}$ crystal. In general, extracting structure factors from arbitrary volumes in the crystal requires rotating the crystal about more than one axis, which requires further development. Achieving that may improve time-resolved crystallography measurements, where a reaction may sweep through the volume of a crystal~\cite{Ramakrishnan:2020,Chapman:2025,Naumov:2015}, or when the absorption of an optical pulse leads to non-uniform excitation of photoactive molecules in a crystal~\cite{Schotte:2003}. It should also improve the analysis of binding in crystals, such as in compound-screening crystallography measurements where ligands diffuse into protein crystals or in the binding of compounds into host framework structures~\cite{Cai:2019,Han:2023}. The union of diffraction and imaging in convergent-beam X-ray diffraction should open many analysis opportunities for a diverse range of molecular, macromolecular and polymeric crystals.

\backmatter




\section*{Acknowledgements}
We would like to thank Tjark Delmas and Julia Maracke (DESY) for engineering and technical support, and Joanne Etheridge (Monash University), Dušan Turk (Josef Stefan Institute), and TJ Lane (DESY) for discussions. We acknowledge DESY (Hamburg, Germany), a member of the Helmholtz Association HGF, for support and for the provision of experimental facilities. Part of this research was carried out at PETRA III. Data was collected using beamline P11 operated by DESY Photon Science. We would like to thank Johanna Hakanpää and her colleagues for assistance during the experiments. Beamtime was allocated for proposals I-20211397, I-20220442, and I-20231166. This work was further supported by the Cluster of Excellence “CUI: Advanced Imaging of Matter” of the Deutsche
Forschungsgemeinschaft (DFG)—EXC 2056—project ID 390715994. L.K. and T.B. acknowledge the support by the Bundesministerium für Forschung, Technologie und Raumfahrt (BMFTR) within the Röntgen-Ångström-Cluster 05K2024 - 2023-06386 DYNAMIX-SP


\section*{Data Availability}
The data that support this study are available from the corresponding authors upon request. 

\section*{Code Availability}
The SPIND indexing software is available under the GNU General Public License from https://github.com/LiuLab-CSRC/SPIND.

\section*{Author Contributions}
The research project was conceived by C.F.L., H.N.C. and S.B. X-ray multilayer Laue lenses were developed by S.B. with input from H.N.C., M.Z., M.P., J.C.W., J.L.D., and H.F. Instrumentation was developed by H.F. with controls software provided by I.D.G.A., D.E., and P.M. Convergent-beam diffraction experiments were carried out by C.F.L., H.N.C., S.B., M.Z., M.P., J.C.W., J.L.D., N.I., W.Z., O.Y., M.Ba., and M.Bu. Analysis was led by C.F.L. with support from M.Z., J.L.D., N.I., W.Z., O.Y., M.Bu., and H.N.C. Si crystal samples were fabricated by M.P. and molecular crystals grown and characterised by D.O., A.H., J.S., and B.K. Structure solution and refinement of vitamin B$_{12}$ was performed by D.O. and L.K.





\bibliography{CBXD-bib}

\end{document}